\pdfoutput=1

\documentclass[a4paper,fleqn]{cas-dc}

\usepackage{microtype}
\usepackage{textcomp}
\usepackage{amssymb}
\usepackage{amsmath}

\usepackage{graphicx}
\graphicspath{{./}}
\usepackage{float}
\usepackage{subfigure}
\usepackage{subcaption}
\usepackage{dcolumn}
\usepackage{bm}
\usepackage{multirow}
\usepackage{caption}
\usepackage{overpic}
\usepackage{booktabs}
\usepackage{color}
\usepackage{array}
\usepackage{breakcites}

\usepackage[numbers,sort&compress]{natbib}
\usepackage{hyperref}
\usepackage[capitalize,nameinlink]{cleveref}
\hypersetup{
	colorlinks=true, 
	linkcolor=cyan, 
	citecolor=cyan, 
	urlcolor=cyan  
}

\begin{document}
	
\let\WriteBookmarks\relax
\def\floatpagepagefraction{1}
\def\textpagefraction{.001}
\shorttitle{Hybrid functional calculation of electrical activity and complexing mechanism of Cu-related defects}
\shortauthors{Xinyu Shi et~al.}

\title [mode = title]{Hybrid functional calculation of electrical activity and complexing mechanism of Cu-related defects} 

\author[1]{Xinyu Shi}
\credit{Writing – original draft, Methodology, Investigation, Formal analysis}
\author[1,2]{Zirui He}
\credit{Conceptualization, Methodology}
\author[1]{An-An Sun}
\credit{Conceptualization, Methodology}
\author[2]{Siqing Shen}
\credit{Formal analysis, Writing – review \& editing}
\author[2]{Yongli Liang}
\cormark[1]
\credit{Conceptualization, Writing – review \& editing, Project administration}
\author[2]{Hao Hu}
\cormark[1]
\credit{Conceptualization, Writing – review \& editing, Project administration}
\author[1,2]{Shang-Peng Gao}[orcid=0000-0003-4877-3363]
\cormark[1]
\credit{Writing – review \& editing, Supervision, Methodology}
\author[2]{Meng Chen}
\credit{Writing – review \& editing, Supervision, Conceptualization}
\cormark[1]
\cortext[1]{Corresponding authors. E-mail address: Yongli.Liang@ast.com.cn (Y. Liang); hhu@ast.com.cn (H. Hu); gaosp@fudan.edu.cn (S. Gao); mchen@ast.com.cn (M. Chen).}

\affiliation[1]{organization={College of Smart Materials and Future Energy, Fudan University},
	postcode={Shanghai 200433},
	country={China}}

\affiliation[aff2]{organization={Shanghai Advanced Silicon Technology Co., Ltd.},
	postcode={Shanghai 201616},
	country={China}}

\begin{graphicalabstract}
\centering
\includegraphics[width=0.9\linewidth]{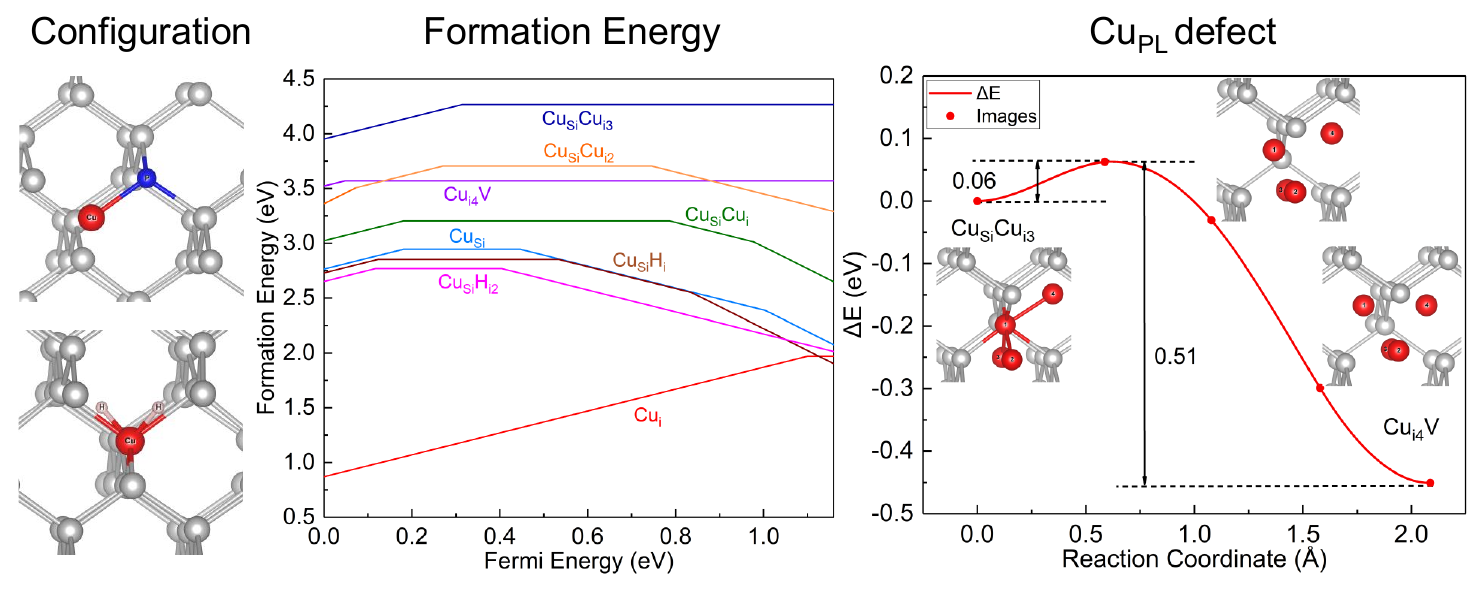}
\label{graphicalabstract}
\end{graphicalabstract}

\begin{highlights}
\item Finite-size corrections are systematically treated for Cu-related defects in Si
\item Configurations and transition levels of Cu-H defects that match experiments are derived
\item $\mathrm{Cu_{i4}V}$ provides an improved description of the $\mathrm{Cu_{PL}}$ center
\end{highlights}


\begin{abstract}
Copper is a detrimental impurity in silicon with high diffusivity and high tendency to precipitate. Interaction between Cu and other defects is essential for understanding the nature of Cu precipitation in silicon. Despite extensive experimental investigations of Cu-related defects in silicon, a comprehensive understanding remains elusive due to limitations of techniques in resolving defect configurations, as well as inconsistencies between theoretical and experimental results regarding transition levels. Moreover, the underlying formation mechanism of the well-known $\mathrm{Cu_{PL}}$ line is still unclear. In this work, configurations, formation energies, and transition levels of Cu-related defects in silicon are calculated using the HSE06 functional and finite-size correction. Defects involved in this study include $\mathrm{Cu_i}$, $\mathrm{Cu_{Si}}$, Cu-B, Cu-P, and Cu-H. A $\mathrm{Cu_{i4}V}$ model is proposed to explain the discrepancies between theory and experiment about $\mathrm{Cu_{PL}}$ defect. Our calculations may provide insight into the electrically active defects and the early states of Cu precipitation in silicon. 
\end{abstract}

\maketitle

\section{Introduction}
Copper is a common but undesirable contaminant in silicon-based devices and solar cells~\cite{claeys2018metal}. It has high diffusivity, strong temperature dependence of solubility, and a tendency to form deep-level recombination centers~\cite{istratov1998intrinsica,istratov2001physics}. It has been reported that Cu may lead to gate oxide failure~\cite{honda1984breakdown,hiramoto1989degradation}, carrier lifetime degradation~\cite{istratov1998electrical}, and light-induced degradation in photovoltaic devices~\cite{lindroos2016review,coletti2011impact}. In addition, Cu contamination may cause spatial uniformity issues and a rising dark current in CMOS image sensors~\cite{russo2017dark}. 

The majority of copper contamination in silicon exists in the form of precipitates at wafer surface or extended defects such as dislocations or grain boundaries~\cite{claeys2018metal}. However, a small but unavoidable fraction of Cu atoms form various electrically active defects within bulk silicon, including isolated defects (i.e., $\mathrm{Cu_i}$ and $\mathrm{Cu_{Si}}$) and defect complexes containing hydrogen~\cite{knack2002copperhydrogen}, carbon~\cite{yarykin2016interstitial}, oxygen~\cite{yarykin2011copperrelated}, dopants~\cite{istratov2001physics}, and other Cu atoms~\cite{lowther2010aggregation,yarykin2024copperricha}. These defects govern the early stages of precipitation due to their strong electrical activity and the high diffusivity of $\mathrm{Cu_i}$. Therefore, understanding the configuration, electronic structure, and formation mechanisms of these small-scale defects is essential for predicting and controlling the precipitation behavior of copper in silicon.

Over the past two decades, extensive computational studies based on density functional theory (DFT) have investigated Cu-related defects in silicon ~\cite{estreicher2004firstprinciples,shirai2005molecular,matsukawa2006gettering,markevich2008radiationinduced,shirai2009new,lowther2010aggregation,wright2016firstprinciples,sharan2017hybridfunctional,vincent2020cu, yarykin2021copper, chen2022firstprinciples}. 
Fundamental properties of isolated Cu defects are relatively well understood. $\mathrm{Cu_i}$ mainly exists in the form of $\mathrm{Cu_i^+}$ in silicon and acts as a fast diffuser with a migration barrier of 0.18 eV~\cite{istratov1998intrinsica}. Deep level transient spectroscopy (DLTS) measurements indicated a $(+/0)$ transition level at $E_\text{c}-0.15~\text{eV}$ ~\cite{istratov1997interstitial}, which has been reproduced by first-principles calculation~\cite{vincent2020cu,lee2022transition}. $\mathrm{Cu_{Si}}$ forms when diffusing $\mathrm{Cu_i}$ occupies a pre-existing vacancy. It introduces three transition levels at $E_\text{v}+0.20~\text{eV}$, $E_\text{v}+0.54~\text{eV}$, and $E_\text{v}+0.97~\text{eV}$ in the band gap~\cite{sharan2017hybridfunctional}.

$\mathrm{Cu_{Si}}$ can form stable complexes with hydrogen atoms, where the hydrogen atoms bind directly to $\mathrm{Cu_{Si}}$ rather than to neighboring silicon atoms. These complexes exhibit distinct electrical behavior depending on the number of $\mathrm{H_i}$ atoms. $\mathrm{Cu_{Si}\text{-}H_{i}}$ and $\mathrm{Cu_{Si}\text{-}H_{i2}}$ introduce levels that are shifted from those of $\mathrm{Cu_{Si}}$, while $\mathrm{Cu_{Si}\text{-}H_{i3}}$ is fully passivated and electrically inactive~\cite{latham2005passivation,yarykin2013deep}.

For Cu-dopant interaction, a database of the calculated binding energies between Cu and various dopants has been established~\cite{matsukawa2006gettering,shirasawa2015useful}. The in-depth formation and dissociation mechanism of Cu-B pair in Si has been modeled via first-principles calculation~\cite{wright2016firstprinciples}. Similarly, the interaction between Cu and oxygen defects has been investigated, showing that diffusing $\mathrm{Cu_i}$ can be trapped by OV pairs and transform into more stable configurations after heat treatment~\cite{west2003copper,yarykin2011copperrelated}.


Despite these efforts, accurate theoretical investigation regarding Cu complex defects, including their configurations and electrical properties, is still required. Current theoretical results about Cu-P interaction lack charge-state-dependent correction. For the case of Cu-H complex, discrepancies persist regarding the exact configurations of hydrogen and positions of transition levels of Cu-H complex~\cite{west2003copper,latham2005passivation,sharan2017hybridfunctional}. Consequently, the calculated binding energies and transition levels so far showed limited agreement with DLTS results~\cite{knack2002copperhydrogen,yarykin2013deep}.


Another unresolved Cu defect in silicon is the $\mathrm{Cu_{PL}}$ center, named after a zero-phonon line at 1.014 eV observed in photoluminescence (PL) studies~\cite{weber1982optical,brotherton1987deepa}. Although a donor level at $E_\text{v}+0.10~\text{eV}$~\cite{brotherton1987deepa} and trigonal symmetry have been reported~\cite{weber1982optical}, its atomic configuration remains unclear. High-resolution PL studies with isotopically pure Si suggested that $\mathrm{Cu_{PL}}$ includes four copper atoms~\cite{steger2008reduction}. Most theoretical results support the $\mathrm{Cu_{Si}\text{-}Cu_{i3}}$ model, but the estimated $(+/0)$ transition level lies above $E_\text{v}+0.30~\text{eV}$~\cite{carvalho2011fourcopper,sharan2017hybridfunctional,vincent2020cu}, which is a significant deviation from DLTS  results~\cite{weber1982optical,istratov1998dissociation}. Another possible structure including four interstitial copper atoms, namely $\mathrm{Cu_{i4}V}$, was calculated to have a $(+/0)$ transition level in between $E_\text{v}+0.07~\text{eV}$ and $E_\text{v}+0.11~\text{eV}$~\cite{fujimura2021revisiting}.

The aim of our work is to investigate Cu-related defects within a unified theoretical framework and to provide insight into their stability and interaction mechanisms. Defects involved in this study are isolated Cu defects, Cu-B, Cu–P, and Cu–H complexes. We also re-examine the possible configurations for $\mathrm{Cu_{PL}}$. Our calculated transition levels, obtained with finite-size corrections, show good consistency with experimental observations.

\section{Methods}
Our calculations were carried out using the {\fontsize{8pt}{10pt}\selectfont VASP} code~\cite{kresse1996efficient}, based on the generalized Kohn-Sham theory and the HSE06 hybrid functional~\cite{heyd2006erratum}. The interactions between valence electrons and ionic cores were treated with the projected-augmented-wave potentials~\cite{blochl1994projector}. Defects were placed in a 64-atom supercell, except for the case of $\mathrm{Cu_{PL}}$, where a 216-atom supercell was used. A $2\times2\times2$ Monkhorst-Pack mesh was used to sample the Brillouin zone~\cite{monkhorst1976special}. All defect geometries were optimized with conjugate gradient algorithm. For simplicity, we use the term ``defect'' to refer to intrinsic defects, impurities, and complexes. 

The calculated equilibrium lattice parameter of Si is 5.434~\textnormal{\AA}, which is very close to the experiment value of 5.431~\textnormal{\AA}~\cite{straumanis1961perfection}. The calculated band gap is 1.16 eV, which is consistent with the experimentally measured band gap of 1.12~eV~\cite{bludau1974temperaturea}. The static dielectric constant calculated by density functional perturbation theory is 12.62.

The formation energy of a specific defect $X$ with charge $q$ is defined as~\cite{freysoldt2014firstprinciples}

\begin{equation}
	E_\text{f}[X^q]=E_\text{t}[X^q]-E_\text{t}[\text{host}]-\sum_{i}{n_{i}\mu_{i}}+qE_\text{F}+E_\text{corr},
\end{equation}
where $E_\text{t}[X^q]$ stands for total energy of a defective supercell with charge $q$, and $E_\text{t}[\text{host}]$ stands for total energy of host supercell. The term $\sum_{i}{n_{i}\mu_{i}}$ represents the energy cost of extra $n_{i}$ atoms with chemical potential $\mu_{i}$ to constitute the defect. The chemical potentials are taken from the most stable phases of the corresponding elements. The Fermi level $E_\text{F}$ is referenced to the valence band maximum (VBM) of host material. Finally, $E_\text{corr}$ stands for charge-state-dependent correction term due to finite-size effects~\cite{freysoldt2011electrostatic}.

The transition level $\varepsilon(q_1/q_2)$ is defined as: 

\begin{equation}
	\varepsilon(q_1/q_2)=\frac{E_\text{f}[X^{q_2}]-E_\text{f}[X^{q_1}]}{q_1-q_2}. 
\end{equation}

Due to the linear dependence of formation energy on the Fermi level, charge states with more electrons (more negative charge states) become energetically more favorable as the Fermi level increases, and vice versa.
 The transition level thus represents the Fermi level where two charge states are thermodynamically equally stable. Note that the configurations of $X^{q_1}$ and $X^{q_2}$ are not essentially the same since each charge state was optimized individually.

For defect complexes, the binding energy of a defect complex A-B is defined as the difference of formation energy between defect complex and the sum of isolated constituents: 
\begin{equation}
	E_\text{b}[AB]=E_\text{f}[A]+E_\text{f}[B]-E_\text{f}[AB].
\end{equation}

The diffusion paths and diffusion barriers were calculated with the climbing image nudged elastic band (CI-NEB) method, implemented in the vtst-tools~\cite{henkelmanClimbingImageNudged2000}. Five images were used, with a spring constant of 5 $\mathrm{eV/\textnormal{\AA}^2}$ between adjacent images. 

\section{Results and discussion}
\subsection{Isolated defects}
Since $\mathrm{Cu_i}$ and $\mathrm{Cu_{Si}}$ have been extensively studied, here we briefly summarize our results and compare them with previous conclusions. For $\mathrm{Cu_i}$, we identify a $(+1/0)$ transition level at $E_\text{c}-0.08~\text{eV}$. Previous DFT results show discrepancy regarding this transition level. A level detected by DLTS at $E_\text{c}-0.15~\text{eV}$ was previously assigned to donor state of $\mathrm{Cu_i}$ and verified by DFT calculations~\cite{istratov1997interstitial,vincent2020cu,lee2022transition}. However, some DFT results suggest that $\mathrm{Cu_i}$ does not introduce level in the band gap~\cite{sharan2017hybridfunctional}. Our calculation shows that $\mathrm{Cu_i}$ is positively  charged in most cases, except in heavily-doped $n$-type silicon where the neutral state becomes stable. 

\begin{figure}[htbp]
	\centering
	\includegraphics[width=0.9\linewidth]{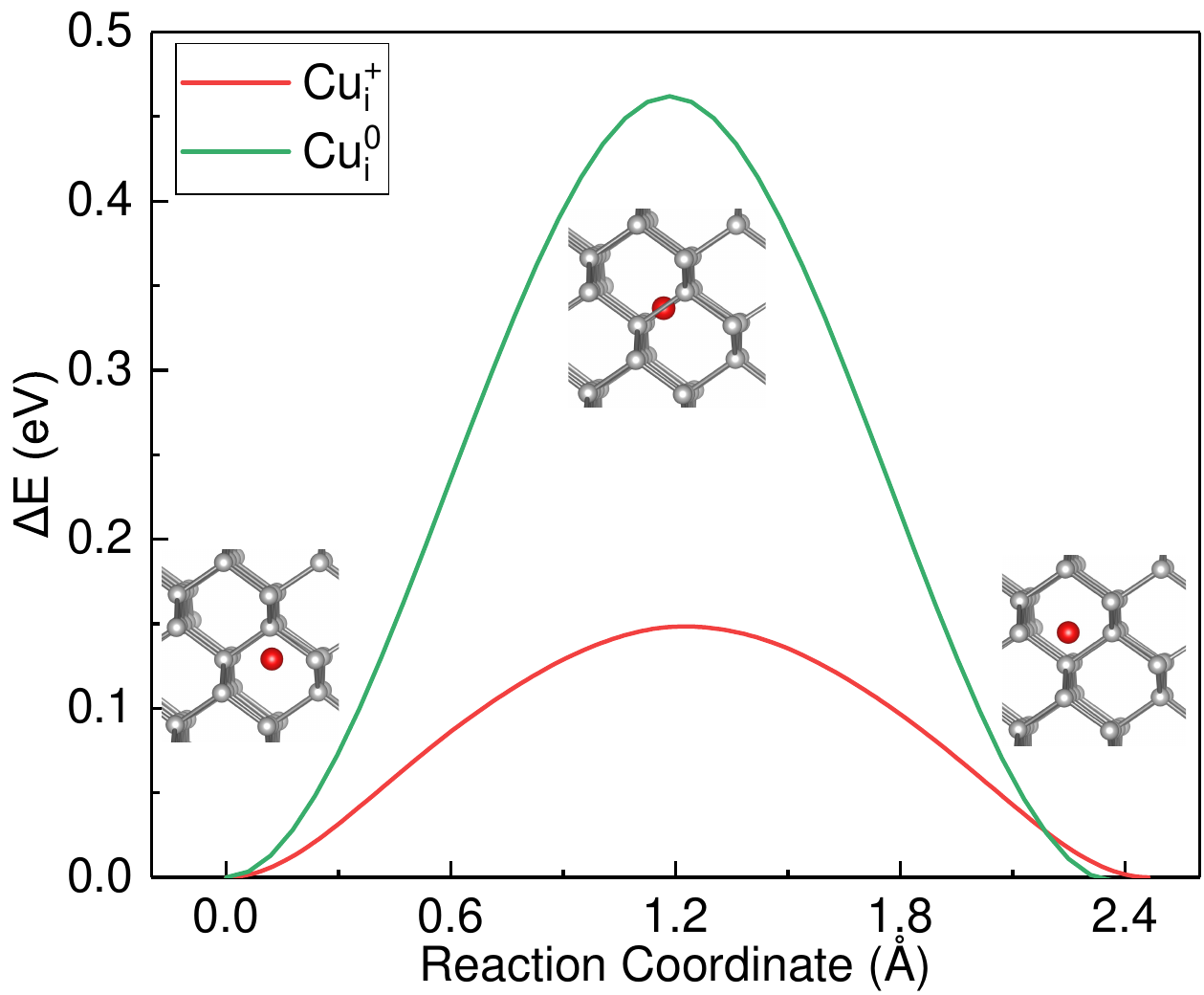}
	\caption{Energy barriers for $\mathrm{Cu_i^+}$ and $\mathrm{Cu_i^0}$ diffusion in silicon. Atomic configurations of $\mathrm{Cu_i}$ at initial state (T-site), saddle point (H-site) and final state (T-site) are shown in the insets.}
	\label{Cui_NEB}
\end{figure}

$\mathrm{Cu_i}$ diffuses via an interstitial mechanism in intrinsic silicon~\cite{istratov1998intrinsica}. Its diffusion pathway initiates from a tetrahedral site (T-site), passes through a hexagonal site (H-site), and terminates at another T-site, along the $\langle111\rangle$ direction. Our NEB calculation with HSE06 functional is shown in~\cref{Cui_NEB}. The calculated diffusion barrier of $\mathrm{Cu_{i}^+}$ is 0.15 eV, whereas that of $\mathrm{Cu_{i}^0}$ is 0.46 eV. Our result for $\mathrm{Cu_{i}^+}$ is consistent with intrinsic diffusion barrier of $0.18\pm0.02~\text{eV}$ obtained by transient ion drifting (TID) experiment~\cite{istratov1998intrinsica}. For comparison, our calculated diffusion barrier with PBE-GGA functional for $\mathrm{Cu_{i}^+}$ is 0.13 eV, which is close to a previous LDA result of 0.11 eV~\cite{matsukawa2007diffusion}. It should be noted that both LDA and GGA functionals tend to underestimate reaction barriers due to the intrinsic approximations in the exchange–correlation functional~\cite{weiske2025adsorption}. 

To date, the diffusion barrier of $\mathrm{Cu_i^0}$ in $n$-type silicon has not been directly quantified by experiments, primarily due to its charge neutrality. Additionally, its relatively high formation energy may further suppress its concentration, thus hindering experimental observation.

\begin{figure}[htbp]
	\centering
	\includegraphics[width=0.9\linewidth]{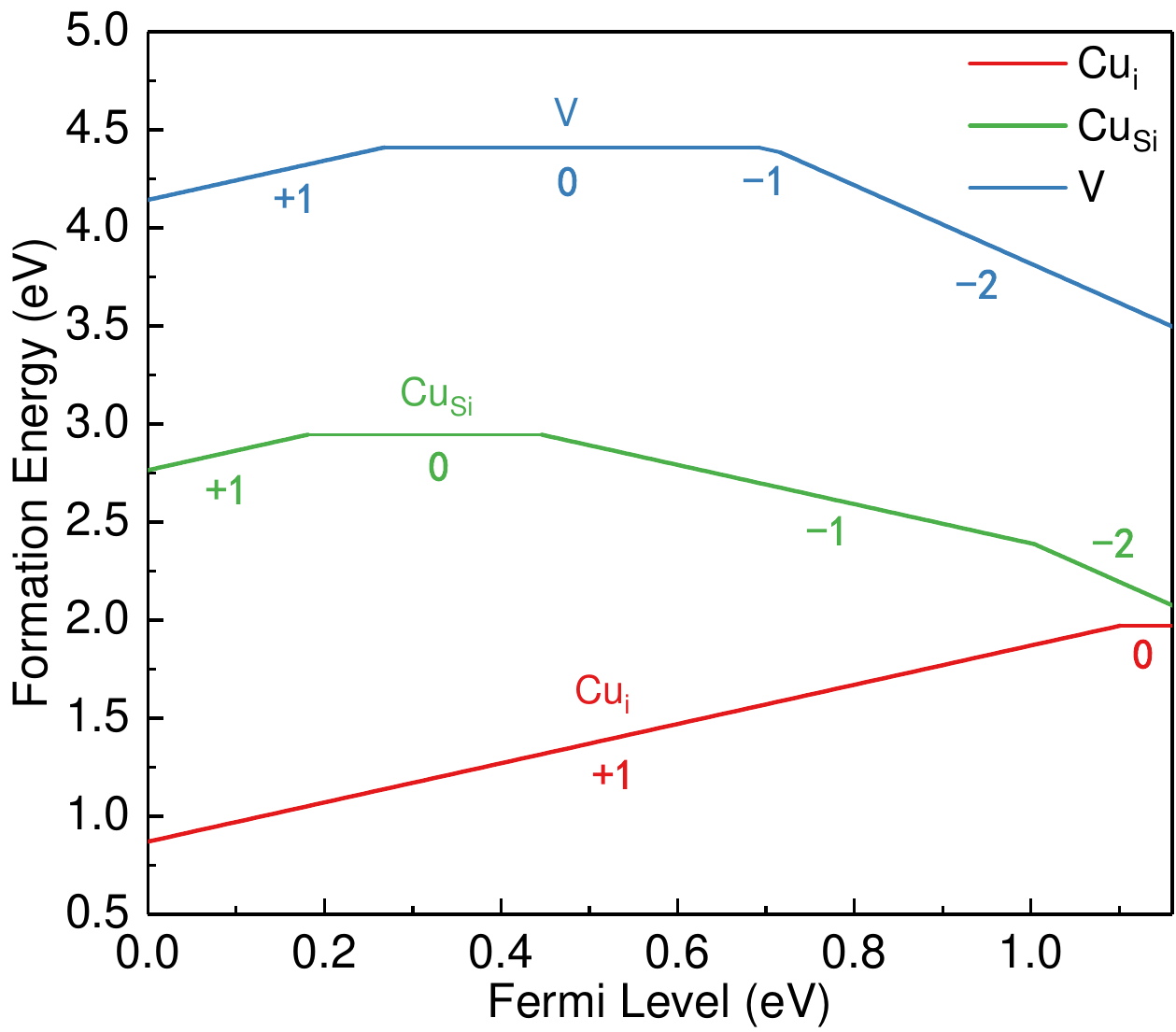}
	\caption{Calculated formation energies of isolated defects, including vacancy, $\mathrm{Cu_i}$, and $\mathrm{Cu_{Si}}$. }
	\label{FormE_Atomic}
\end{figure}

According to theoretical prediction, $\mathrm{Cu_{Si}}$ is subject to Jahn-Teller distortion, leading to a symmetry reduction~\cite{drabold2007theory}. However, previous DFT results show varied results on the exact symmetries of $\mathrm{Cu_{Si}}$ with different charge states~\cite{latham2005passivation,latham2006electronica,sharan2017hybridfunctional}. Our calculations find that structural relaxations with different initial symmetries yield slightly different geometries, but the corresponding formation energy differences are consistently smaller than 0.01~eV, which is well below the uncertainty caused by the finite-size correction.
The near-degeneracy indicates that distortion around $\mathrm{Cu_{Si}}$ is exceptionally subtle, and the potential energy surface (PES) of $\mathrm{Cu_{Si}}$ is so gentle that configurations with multiple symmetries correspond to nearly equivalent local minima. Consequently, these small differences have a negligible impact on the calculated charge-transition levels and binding energies, and do not affect our main conclusions.

Transition levels of $\mathrm{Cu_{Si}}$ in this work and in previous works are listed in~\cref{Cusi_Level}. Our results generally agree with previous DLTS results~\cite{brotherton1987deepa,knack2002copperhydrogen,yarykin2011copperrelated}. Regarding the $(0/-1)$ level of $\mathrm{Cu_{Si}}$, some previous DFT results show discrepancies with experiments, whereas the deviation in this work is relatively small. $\mathrm{Cu_{Si}}$ has +1, 0, $-1$, and $-2$ charge states in the band gap, with spin state of $S$=0, $1/2$, 0, and $1/2$ respectively. Therefore, in low to moderately doped silicon samples, $\mathrm{Cu_{Si}}$ is typically in the 0 or $-$1 charge state, depending on the exact temperature and doping level. In heavily-doped $p$-type silicon, $\mathrm{Cu_{Si}}$ is in the $+$1 charge state. In heavily-doped $n$-type silicon, $\mathrm{Cu_{Si}}$ is predominantly in the $-2$ charge state. 
\begin{table}
	\centering
	\renewcommand\arraystretch{1.2}
	\setlength{\tabcolsep}{8pt}
	\caption{Transition levels $\mathrm{Cu_{Si}}$ in this work and previous researches. All values are presented in eV relative to $E_\text{v}$. An asterisk (*) means that the acceptor level is originally related to $E_\text{c}$, and recalculated to $E_\text{v}$, assuming a band gap of 1.12 eV.}
	\begin{tabular}{p{1.5cm} p{0.4cm} p{0.4cm} p{0.4cm} p{0.4cm} p{0.4cm} p{0.4cm} p{0.4cm} }
		\toprule
		Transition level & t.w. & & calc. & & & exp. &\\
		\midrule
		$(+1/0)$  & 0.18 &  0.17$^a$ & 0.20$^b$ & 0.20$^c$ & 0.23$^d$ & 0.21$^e$  & 0.22$^f$\\
		$(0/-1)$  & 0.45 & 0.62$^a$ & 0.63$^b$ & 0.40$^c$ & 0.43*$^d$ & 0.48$^e$ & 0.43$^f$\\
		$(-1/-2)$ & 1.00 & 0.96$^a$ & 0.97$^b$ & 0.94$^c$ & 0.94*$^d$ & 0.95$^e$ & \\
		\bottomrule
		\multicolumn{8}{p{0.45\textwidth}}{$^{\mathrm{a}}$ Calculated with marker method in a 64-atom supercell. See Ref.~\cite{latham2005passivation}} \\
		\multicolumn{8}{p{0.45\textwidth}}{$^{\mathrm{b}}$ Calculated with finite-size correction in a 64-atom supercell. See Ref.~\cite{sharan2017hybridfunctional}} \\
		\multicolumn{8}{p{0.45\textwidth}}{$^{\mathrm{c}}$ Calculated with finite-size correction in a 216-atom supercell. See Ref.~\cite{vincent2020cu}.}\\
		\multicolumn{8}{p{0.45\textwidth}}{$^{\mathrm{d}}$ Experimental results by DLTS. See Ref.~\cite{brotherton1987deepa}} \\
		\multicolumn{8}{p{0.45\textwidth}}{$^{\mathrm{e}}$ Experimental results by DLTS. See Ref.~\cite{knack2002copperhydrogen}} \\
		\multicolumn{8}{p{0.45\textwidth}}{$^{\mathrm{f}}$ Experimental results by DLTS. See Ref.~\cite{yarykin2011copperrelated}}
	\end{tabular}
	\label{Cusi_Level}
\end{table}

~\cref{FormE_Atomic} shows the formation energy of vacancy, $\mathrm{Cu_{i}}$, and $\mathrm{Cu_{Si}}$. Over a wide range of Fermi energy, $\mathrm{Cu_{i}^+}$ has the lowest formation energy, and in heavily-doped $n$-type silicon, formation energy of $\mathrm{Cu_{Si}^+}$ reduces to slightly above 2 eV. Moreover, $\mathrm{Cu_{Si}}$ forms by $\mathrm{Cu_i}$ filling a pre‑existing vacancy ($E_\text{b}=2.51$~eV for reaction $\mathrm{Cu_i^+ + V^0 \rightarrow Cu_{Si}^+}$~), rather than by $\mathrm{Cu_i}$ directly displacing a Si atom (which is highly endothermic, $E_\text{b}=-6.09$~eV for reaction $\mathrm{Cu_i^+ \rightarrow Cu_{Si}^+ + Si_i^0}$~).

Our calculation also reveals the microscopic origin of the phenomenon that Cu exists primarily as interstitials in $p$-type silicon and tends to form $\mathrm{Cu_{Si}}$ or electrically inactive silicide in heavily-doped $n$-type silicon~\cite{istratov2001physics}. In $p$-type silicon, $\mathrm{Cu_i}$ is favored due to its low formation energy. In moderately-doped $n$-type silicon, both interstitial and substitutional forms can coexist. In heavily-doped $n$-type silicon, the reduced formation energy of vacancies provides more sites for $\mathrm{Cu_{Si}}$ formation and subsequent generation of electrically inactive silicides.

\subsection{Cu-B and Cu-P complex}
Boron and phosphorus are the most widely used $p$-type and $n$-type dopants in silicon. They are known to influence transition metal impurities through the 
``ion pairing" effect~\cite{ozaki2019gettering}. Since boron and phosphorus are both shallow dopants in silicon~\cite{bystrom2024nonlocal}, here we assume them to be completely ionized. 

\begin{figure}
	\subfigcapskip=-1pt 
	\centering
	\subfigure{
		\put(-15,130){\textbf{(a)}}
		\label{CuBstructure}
		\includegraphics[width=0.6\linewidth]{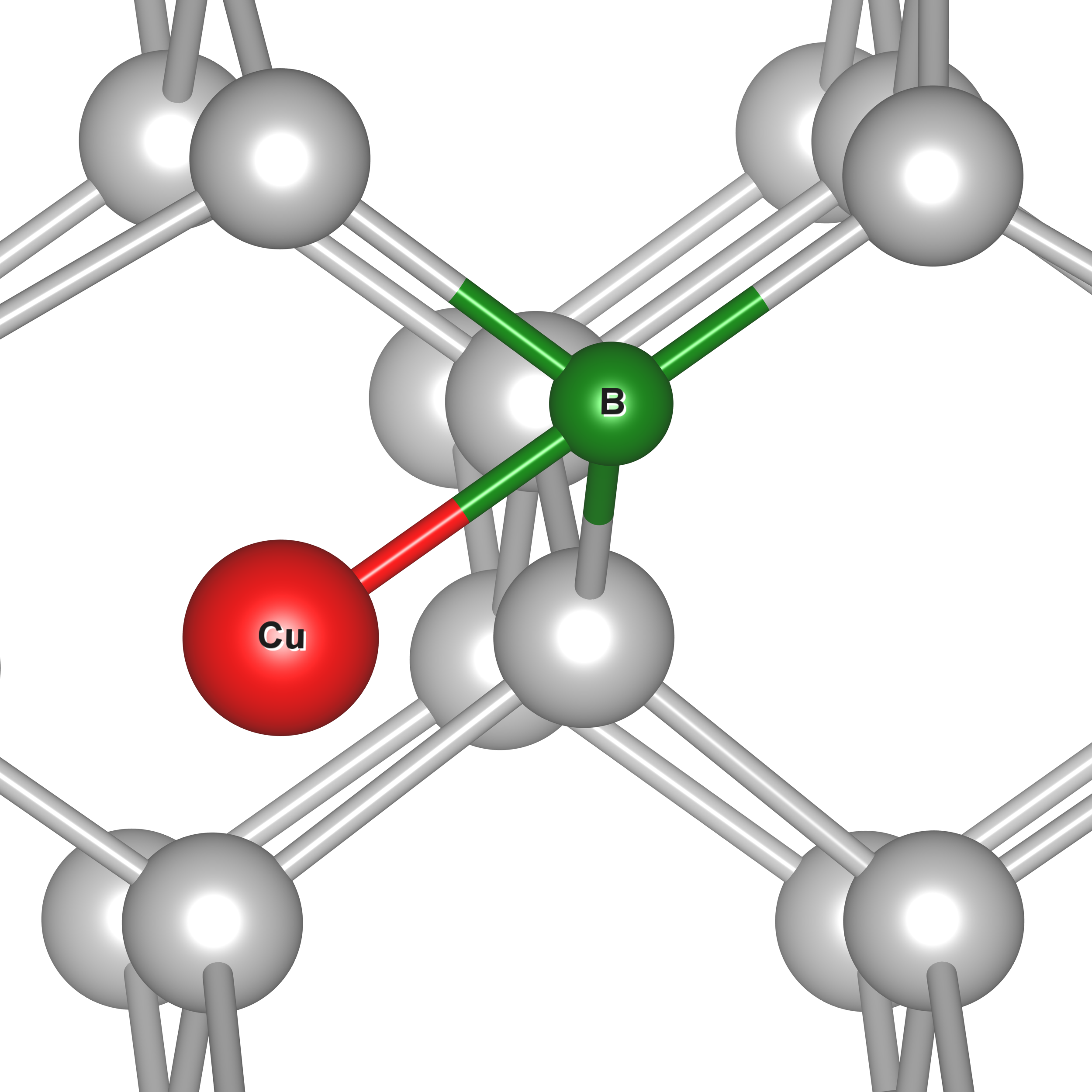}}
	\subfigure{
		\put(-15,130){\textbf{(b)}}
		\label{CuPstructure}
		\includegraphics[width=0.6\linewidth]{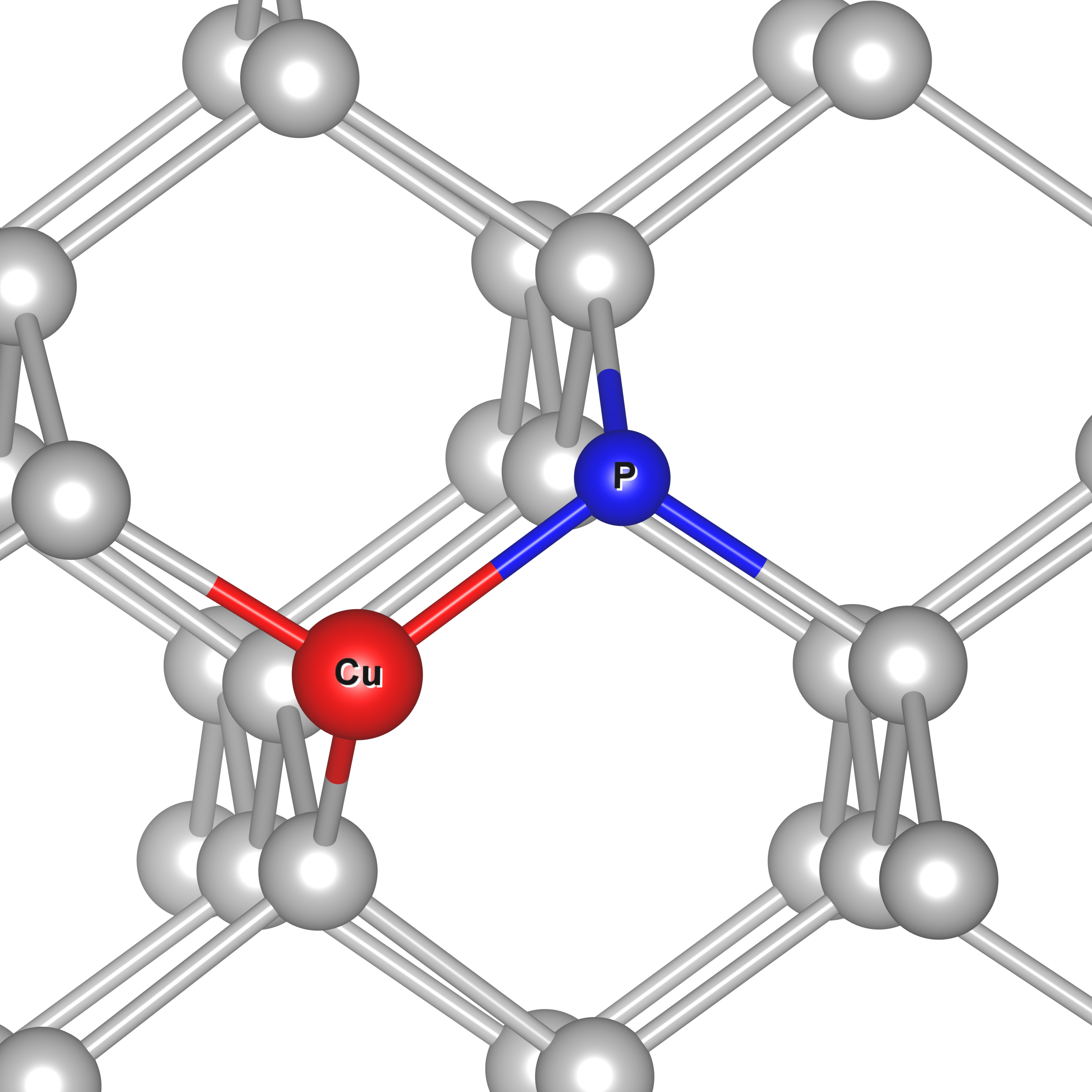}}
	\caption{Configurations of (a) $\mathrm{Cu_{i}\text{-}B_{Si}}$ and (b) $\mathrm{Cu_{Si}\text{-}P_{Si}}$. The red atom is copper, the green atom is boron, and the blue atom is phosphorus.}
\end{figure}

In $p$-type silicon, Cu mainly exists as $\mathrm{Cu_{i}^+}$. $\mathrm{Cu_{i}^+}$ and $\mathrm{B_{Si}}$ form a defect complex with $C_{3v}$ symmetry, as shown in~\cref{CuBstructure}. The binding energy between $\mathrm{Cu_i^+}$ and $\mathrm{B_{Si}^-}$ is $E_b=0.44~\text{eV}$. The dissociation energy $E_\text{diss}$, which is defined as $E_\text{diss}=E_\text{b}+E_\text{d}$, is 0.59~eV, in agreement with previous TID result of 0.61~eV~\cite{wright2016firstprinciples}. This relatively small dissociation energy indicates that the binding between $\mathrm{Cu_{i}^+}$ and $\mathrm{B_{Si}}$ is quite weak. As a result, the lifetime of $\mathrm{Cu_{i}\text{-}B_{Si}}$ is less than 1~ms~\cite{istratov1998intrinsica}, and dissociated $\mathrm{Cu_i}$ can readily diffuse to more stable trapping sites like extended defects or wafer surface.  

In $n$-type silicon, both $\mathrm{Cu_{Si}}$ and $\mathrm{Cu_i}$ can exist as discussed above. The binding energy of $\mathrm{Cu_{i}^+}$ and $\mathrm{P_{Si}^+}$ is negative due to a repulsive Coulomb force. In contrast, the binding energy of $\mathrm{Cu_{Si}}$ and $\mathrm{P_{Si}^+}$ is larger than 1 eV, as shown in~\cref{Eb_CuBP} and~\cref{CuPstructure}. For comparison, previous DFT results based on neutral defects $\mathrm{Cu_{Si}\text{-}P_{Si}}$ gave a binding energy of only 0.42~eV~\cite{shirasawa2015useful}. 

Moreover, as the Fermi level rises and $\mathrm{Cu_{Si}}$ has more negative charges, the binding energy between $\mathrm{Cu_{Si}}$ and $\mathrm{P_{Si}}$ increases. This trend is consistent with the gettering model proposed by Ozaki et~al~\cite{ozaki2019gettering}, where the dominant gettering reaction in $n$-type silicon is the binding between $\mathrm{Cu_{Si}^{-3}}$ and $\mathrm{P_{Si}}$. In heavily doped $n$-type silicon, formation energy of $\mathrm{Cu_{Si}^{-3}}$ is slightly larger than $\mathrm{Cu_{Si}^{-2}}$, making the transition thermodynamically feasible at elevated temperatures. Consequently, $\mathrm{Cu_{Si}^{-2}}$ can trap an additional electron to form $\mathrm{Cu_{Si}^{-3}}$, which exhibits the strongest binding with $\mathrm{P_{Si}^+}$ among all Cu-P reactions listed in~\cref{Eb_CuBP}, thereby driving the gettering process.

We conclude that phosphorus has a distinct gettering mechanism compared with boron. In heavily phosphorus-doped silicon, Cu can be effectively trapped as $\mathrm{Cu_{Si}\text{-}P_{Si}}$ pair, or forming precipitates at the backside $\mathrm{N^+}$ area~\cite{yakimov2019metal}.

\begin{table}
	\centering
	\renewcommand\arraystretch{1.5}
	\caption{The calculated binding energy ($E_\text{b}$) between $\mathrm{Cu_{Si}}$ and $\mathrm{P_{Si}}$ at various charges.}
	\begin{tabular}{ccc}
		\toprule\rule{0pt}{8pt}
		Reaction   & $E_\text{b}$~/eV       \\ 
		\toprule\rule{0pt}{8pt}
		$\mathrm{Cu_{i}^+ + P_{Si}^+ }$~$\rightarrow$~$\mathrm{{Cu_{i}\text{-}P_{Si}}^{+2}}$ & $-$0.49  \\   
		
		$\mathrm{Cu_{Si}^{-1} + P_{Si}^+ }$~$\rightarrow$~$\mathrm{{Cu_{Si}\text{-}P_{Si}}^{0}}$  & 1.25 \\
		
		$\mathrm{Cu_{Si}^{-2} + P_{Si}^+ }$~$\rightarrow$~$\mathrm{{Cu_{Si}\text{-}P_{Si}}^{-1}}$  & 1.44 \\
		
		$\mathrm{Cu_{Si}^{-3} + P_{Si}^+ }$~$\rightarrow$~$\mathrm{{Cu_{Si}\text{-}P_{Si}}^{-2}}$  & 1.69 \\
		\bottomrule\rule{0pt}{8pt}
	\end{tabular}
	\label{Eb_CuBP}
\end{table}

\subsection{Cu-H complex}
Hydrogen is a ubiquitous impurity that can be introduced during several key semiconductor fabrication processes, including crystal growth~\cite{hara1993hydrogen}, ion implantation~\cite{duo2001evolution}, post-deposition annealing~\cite{jiang2003hydrogenation}, and wet-chemical etching~\cite{nickel2023hydrogen}. Hydrogen can play various roles in crystalline Si, with the best known being the passivation of shallow dopants~\cite{vandewalle1989theory}. Hydrogen also activates electrically inactive impurities, passivates various kinds of deep-level defects, and creates its own electrically and optically active centers~\cite{west2003copper}.

Isolated $\mathrm{H_i}$ has only a single donor state and a single acceptor state in the band gap, with a $(+1/-1)$ transition level at $E_\text{v}+0.81~\text{eV}$. $\mathrm{H_{i}^-}$ and $\mathrm{H_i^0}$ prefer a linear Si-H-Si structure, with a Si-H bond length of 1.59 \textnormal{\AA}. $\mathrm{H_{i}^+}$ prefers an anti-bond site with a Si-H bond of 1.75 \textnormal{\AA} along $\mathrm\langle111\rangle$ direction.

\begin{figure}[htbp]
	\subfigcapskip=-2pt 
	\centering
	\subfigure{
		\put(-10,90){\textbf{(a)}}
		\centering
		\label{CusiHi_C2v}
		\includegraphics[width=0.4\linewidth]{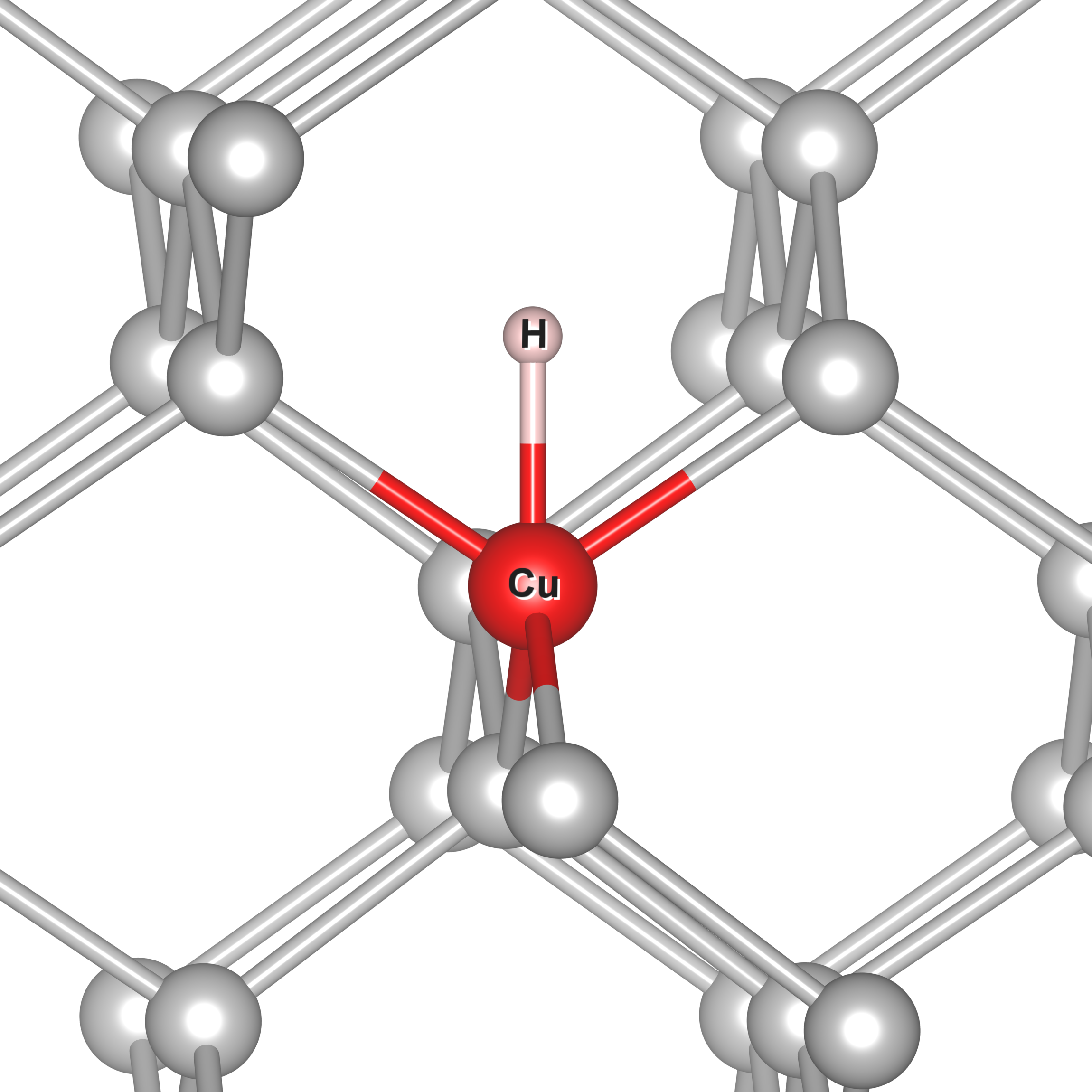}}
	\qquad
	\subfigure{
		\put(-10,90){\textbf{(b)}}
		\label{CusiHi_Cs}
		\includegraphics[width=0.4\linewidth]{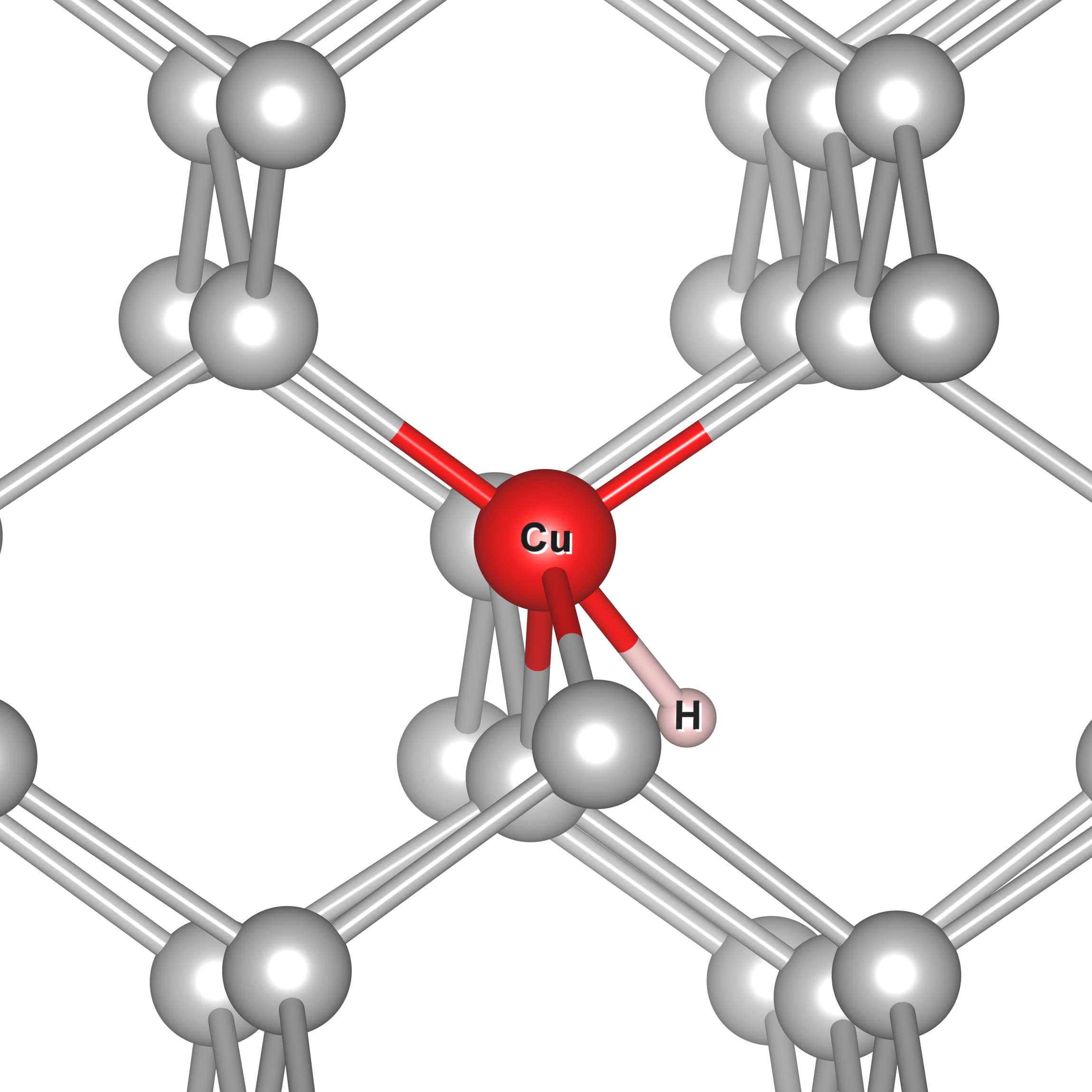}}
	\qquad 
	\subfigure{
		\put(-10,90){\textbf{(c)}}
		\label{CusiHi2_C2}
		\includegraphics[width=0.4\linewidth]{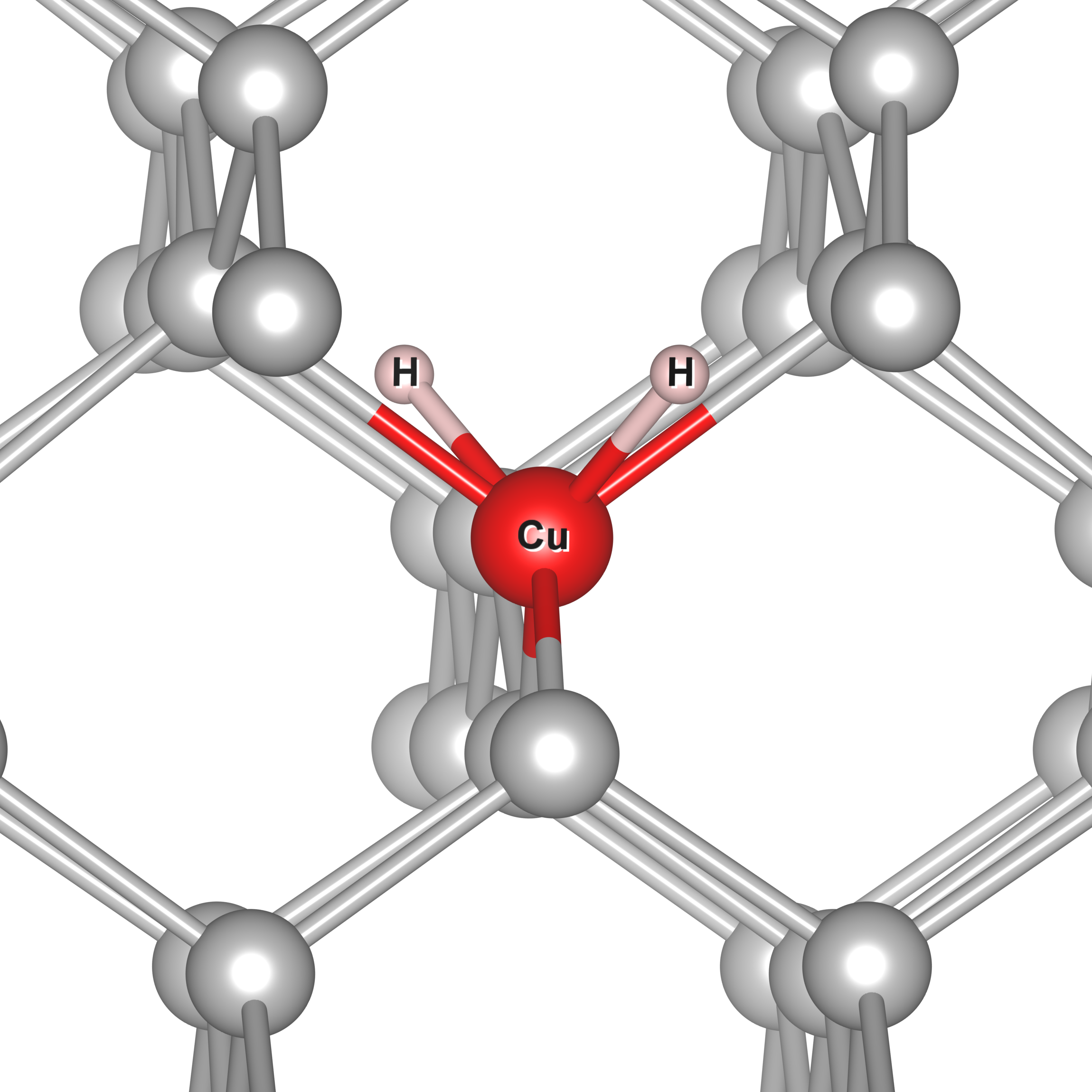}}
	\qquad
	\subfigure{
		\put(-10,90){\textbf{(d)}}
		\label{CusiHi2_Cs}
		\includegraphics[width=0.4\linewidth]{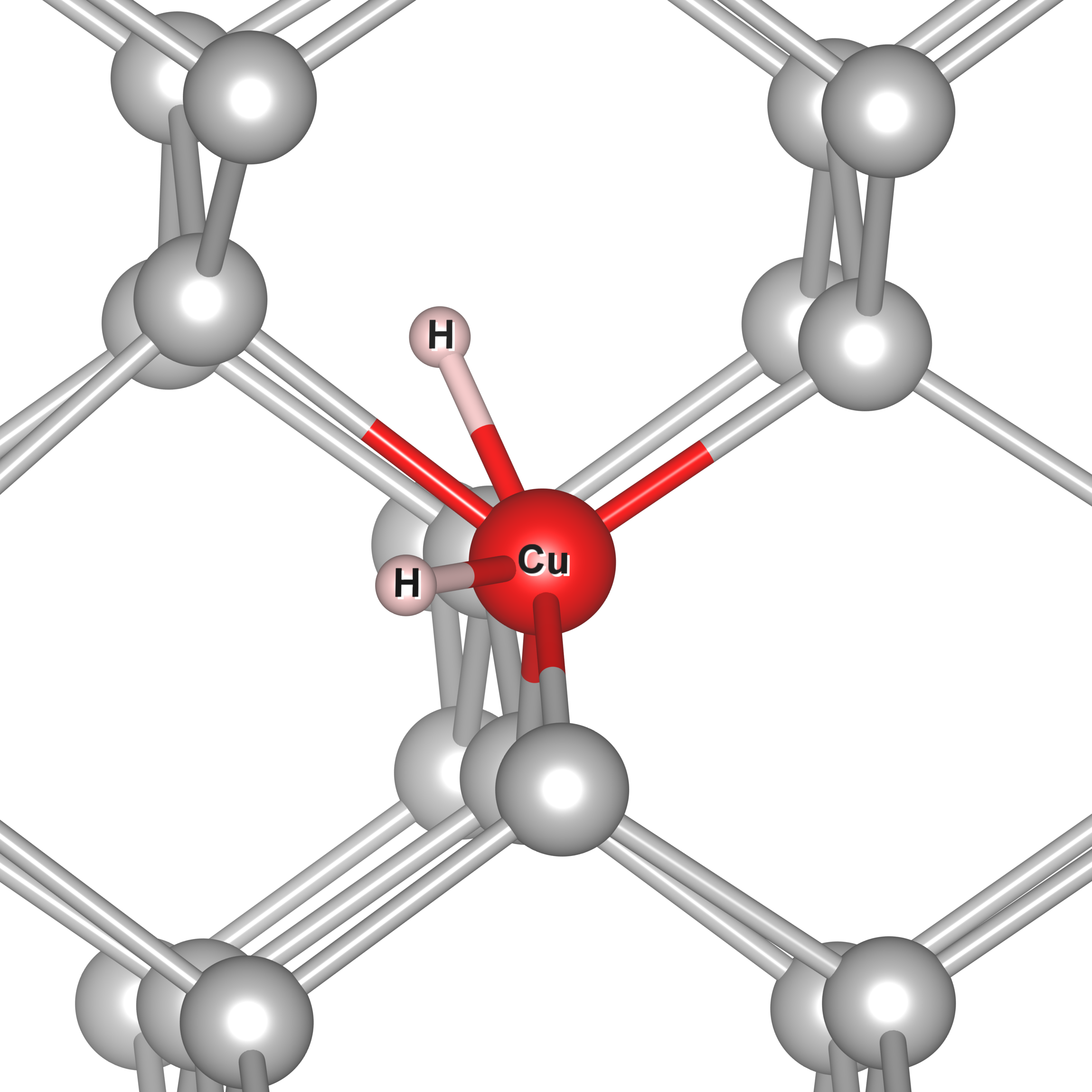}}
	\subfigure{
		\put(-10,185){\textbf{(e)}}
		\label{FormE_CuH}
		\includegraphics[width=0.9\linewidth]{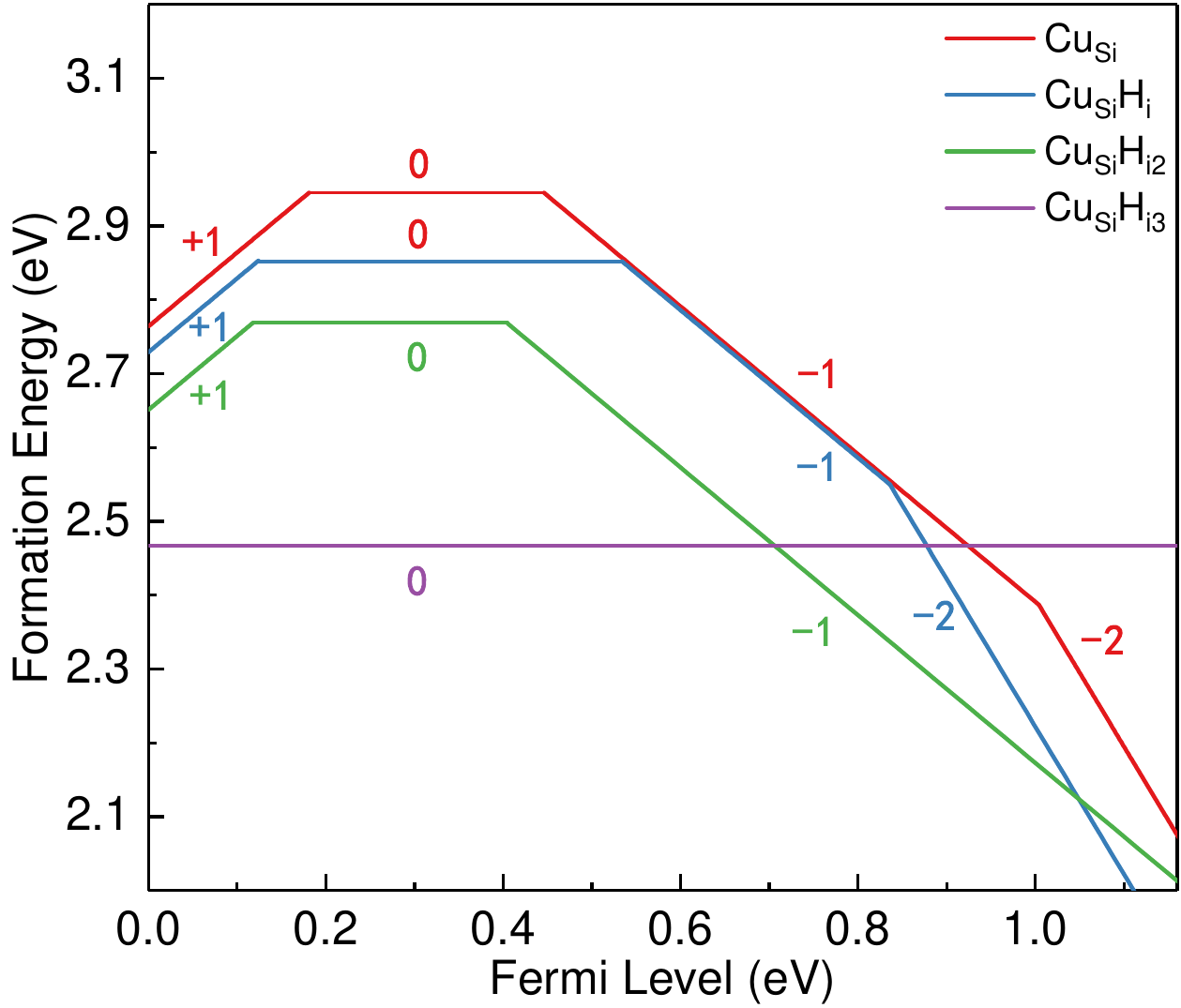}}
	\caption{Configurations of $\mathrm{Cu\text{-}H}$ complex. The vertical and horizontal rightward direction corresponds to $[001]$ and $[110]$ direction separatedly. (a) $\mathrm{Cu_{Si}\text{-}H_{i}}$ with $C_{2v}$ symmetry. (b) $\mathrm{Cu_{Si}\text{-}H_{i}}$ with $C_s$ symmetry.
		(c) $\mathrm{Cu_{Si}\text{-}H_{i2}}$ with $C_2$ symmetry.
		(d) $\mathrm{Cu_{Si}\text{-}H_{i2}}$ with $C_s$ symmetry. (e) Calculated formation energies of $\mathrm{Cu_{Si}}$ and $\mathrm{Cu_{Si}\text{-}H_{in}}$ (n=1, 2, 3) defects.} 
\end{figure}

Cu is known to interact with $\mathrm{H_{i}}$, which modifies the electrical activity of $\mathrm{Cu_{Si}}$~\cite{west2003copper,kaniewski1997hydrogenation}. In contrast to many other transition metal impurities (e.g., Ni, Ag), $\mathrm{H_{i}}$ binds directly with $\mathrm{Cu_{Si}}$ rather than to neighboring silicon atoms~\cite{estreicher2004firstprinciples}. To determine the most stable configurations, we examine three initial structures with Cu-H bond along $\mathrm\langle111\rangle$, $\langle110\rangle$, and $\langle100\rangle$ directions, corresponding to ${C_{3v}}$, ${C_{2v}}$, and ${C_{s}}$ symmmetries, respectively. The results show that energetically favorable configurations of $\mathrm{Cu_{Si}\text{-}H_{i}}$ are charge-dependent. At +1, 0, and $-1$ charge states, $\mathrm{Cu_{Si}\text{-}H_{i}}$ adopts ${C_s}$ symmetry, and the formation energy of ${C_{2v}}$ configuration is 0.2 eV higher. At $-2$ charge state, ${C_{2v}}$ and ${C_s}$ structures are energetically very close. The defect configurations are shown in~\cref{CusiHi_C2v} and~\cref{CusiHi_Cs}. Previous theoretical studies have reported different conclusions on the configurations of Cu-H complex. Sharan et~al. ~\cite{sharan2017hybridfunctional} showed that $\mathrm{Cu_{Si}\text{-}H_{i}}$ has ${C_{2v}}$ symmetry across all states, whereas Latham et~al.~\cite{latham2005passivation} proposed that $\mathrm{Cu_{Si}\text{-}H_{i}}$ has ${C_{2v}}$ symmetry at $-2$ and $-1$ charge states, and either ${C_s}$ or ${C_1}$ at neutral and +1 charge state. Our calculated results are in better agreement with latter one.

Using the lowest-energy configuration at each charge state, we derive three transition levels for the $\mathrm{Cu_{Si}\text{-}H_{i}}$ complex: a $(+1/0)$ donor level at $E_\text{v}+0.12~\text{eV}$, a $(0/-1)$ acceptor level at $E_\text{v}+0.53~\text{eV}$, and a $(-1/-2)$ double acceptor level at $E_\text{v}+0.83~\text{eV}$. These results are in good agreement with the three levels at $E_\text{v}+0.10~\text{eV}$, $E_\text{v}+0.56~\text{eV}$, and $E_\text{v}+0.81~\text{eV}$ obtained for  $\mathrm{Cu_{Si}\text{-}H_{i}}$ complex in previous DLTS measurements~\cite{knack2004copperrelated}.

Based on the configurations above, we consider adding one and two more hydrogen atoms. For $\mathrm{Cu_{Si}\text{-}H_{i2}}$, we find two configurations with $C_2$ and $C_s$ symmetries (See~\cref{CusiHi2_C2} and ~\cref{CusiHi2_Cs}). Difference in formation energy of these two configurations are smaller than 0.10~eV at the +1 and 0 charge states. At $-1$ and $-2$ charge states, formation energy of $C_2$ structure is about 0.11~eV lower. For the two configurations above, we only find a $(+1/0)$ donor level at $E_\text{v}+0.12~\text{eV}$ and a $(0/-1)$ level at $E_\text{v}+0.41~\text{eV}$. Our results are in better agreement with DLTS results reported in Ref.~\cite{nikolai2011formation}, where $\mathrm{Cu_{Si}\text{-}H_{i2}}$ exhibits levels at $E_\text{v}+0.19~\text{eV}$ for $(+1/0)$ transition and $E_\text{v}+0.46~\text{eV}$ for $(0/-1)$ transition. We do not identify the $(-1/-2)$ level at $E_\text{c}-0.25~\text{eV}$ reported in Ref.~\cite{knack2004copperrelated}.

For $\mathrm{Cu_{Si}\text{-}H_{i3}}$, the most energetically stable configuration has only $C_1$ symmetry. Configurations with a higher symmetry, (e.g., $C_2$ and $C_s$) have formation energies 0.2 to 0.3 eV higher at neutral charge state. No transition level is found, which is in accordance with previous DLTS experiments~\cite{yarykin2011copperrelated}. 

\begin{figure}[htbp]
	\centering
	\includegraphics[width=0.9\linewidth]{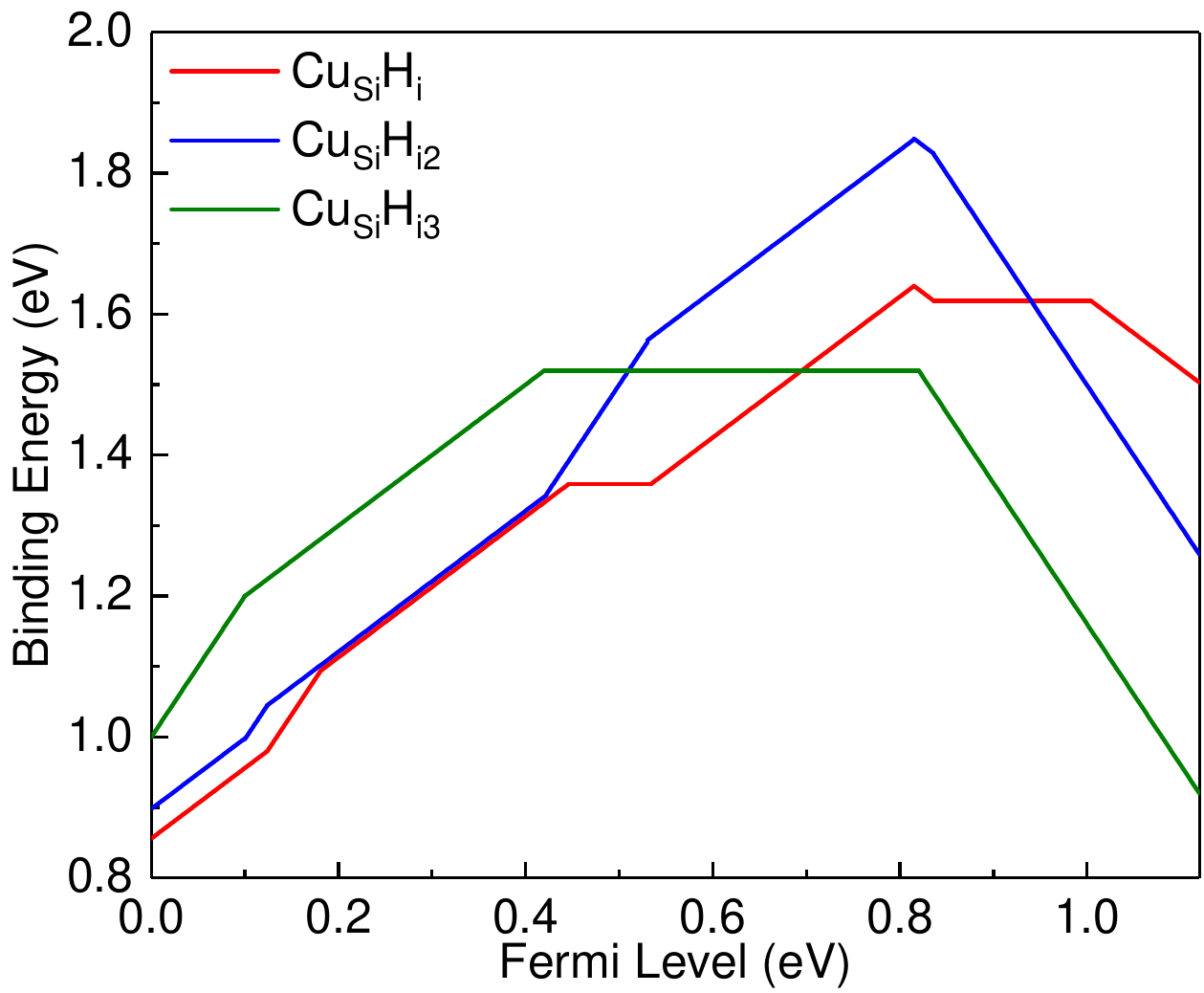}
	\caption{The calculated binding energy ($E_\text{b}$) of Cu-H complex. The energy needed to absorb or release electrons from the environment (i.e., Si) is included in $E_\text{b}$.}
	\label{Eb_CusiHi}
\end{figure}

Finally, we discuss the formation mechanism of $\mathrm{Cu_{Si}\text{-}H_{i3}}$ complex. ~\cref{Eb_CusiHi} shows the calculated binding energy of $\mathrm{Cu_{Si}\text{-}H_{in}}$ (n=1, 2, 3) at various charge states. Although the binding energy of $\mathrm{Cu_{Si}\text{-}H_{i}}$ is higher in $n$-type silicon than in $p$-type silicon, indicating stronger thermodynamic stability, the binding reaction is kinetically hindered by Coulomb repulsion between negatively charged reactants. Therefore, Cu–H complex formation is expected to occur more readily in p-type silicon. $\mathrm{Cu_{Si}\text{-}H_{i2}}$ is more likely to form if the Fermi level falls between 0.53 eV and 0.81 eV. The binding reaction releases an electron to silicon environment, so it would happen more easily in lightly-doped $p$-type silicon. The most energetically favorable pathway to form $\mathrm{Cu_{Si}\text{-}H_{i3}}$ is the reaction between $\mathrm{Cu_{Si}\text{-}H_{i2}^-}$ and $\mathrm{H_i^+}$, which also happens at Fermi level in between 0.42 eV to 0.81 eV. At the same range of Fermi level, the binding energy of a fourth $\mathrm{H_i}$ is smaller than 0.50 eV, which is small compared to that of the first three $\mathrm{H_i}$ atoms. A fourth $\mathrm{H_i}$ would force Cu into an eightfold-coordinated configuration. However, with no empty states available to accommodate valence electron of the fourth $\mathrm{H_i}$, the $\mathrm{Cu_{Si}\text{-}H_{i4}}$ complex becomes energetically unstable. 

Formation energies of $\mathrm{Cu_{Si}}$ and Cu-H complexes are shown in~\cref{FormE_CuH}. Our calculations confirm that up to three H atoms are required to passivate all the levels of $\mathrm{Cu_{Si}}$. Meanwhile, as more $\mathrm{H_i}$ atoms are incorporated, formation energy of Cu-H complex decreases. This trend indicates that hydrogen effectively stabilizes the Cu-related defect. Upon bonding with hydrogen, bonding of $\mathrm{Cu_{Si}}$ shifts from weaker Cu-Si bonds to stronger Cu-H bonds. Moreover, electrons previously trapped at gap states are able to relax into the newly formed, lower-energy bonding states. Consequently, the energy released during this electronic transition is reflected in the decreasing formation energy in the $\mathrm{Cu_{Si}\text{-}H_{in}}$ defect.

It should be noted that previous experiments about $\mathrm{Cu\text{-}H}$ interaction were usually performed in $p$-type Si at a doping concentration of $\mathrm{1\times10^{15}~cm^{-3}}$ ~\cite{yarykin2013deep,knack2002copperhydrogen}. This corresponds to a Fermi level at $E_\text{v}+0.32~\text{eV}$ where $\mathrm{Cu_{Si}\text{-}H_{in}}$ is neutral according to~\cref{FormE_CuH}. Under these conditions, continuous binding of $\mathrm{H_i}$ to $\mathrm{Cu_{Si}}$ requires capturing electrons from the host, which may lead to minority lifetime reduction.

\subsection{$\mathrm{Cu_{PL}}$ defect}
As is mentioned above, two possible defect complexes, namely $\mathrm{Cu_{Si}\text{-}Cu_{i3}}$ and $\mathrm{Cu_{i4}V}$, were proposed to explain the unresolved PL line at 1.014 eV in Cu-contaminated silicon samples~\cite{weber1982optical,brotherton1987deepa}. Here we re-examine both structures with the same scheme used in this work. 

\begin{figure}[!t]
	\subfigcapskip=-2pt 
	\centering
	\subfigure{
		\put(0,175){\textbf{(a)}}
		\label{CuPL_NEB}
		\includegraphics[width=0.95\linewidth]{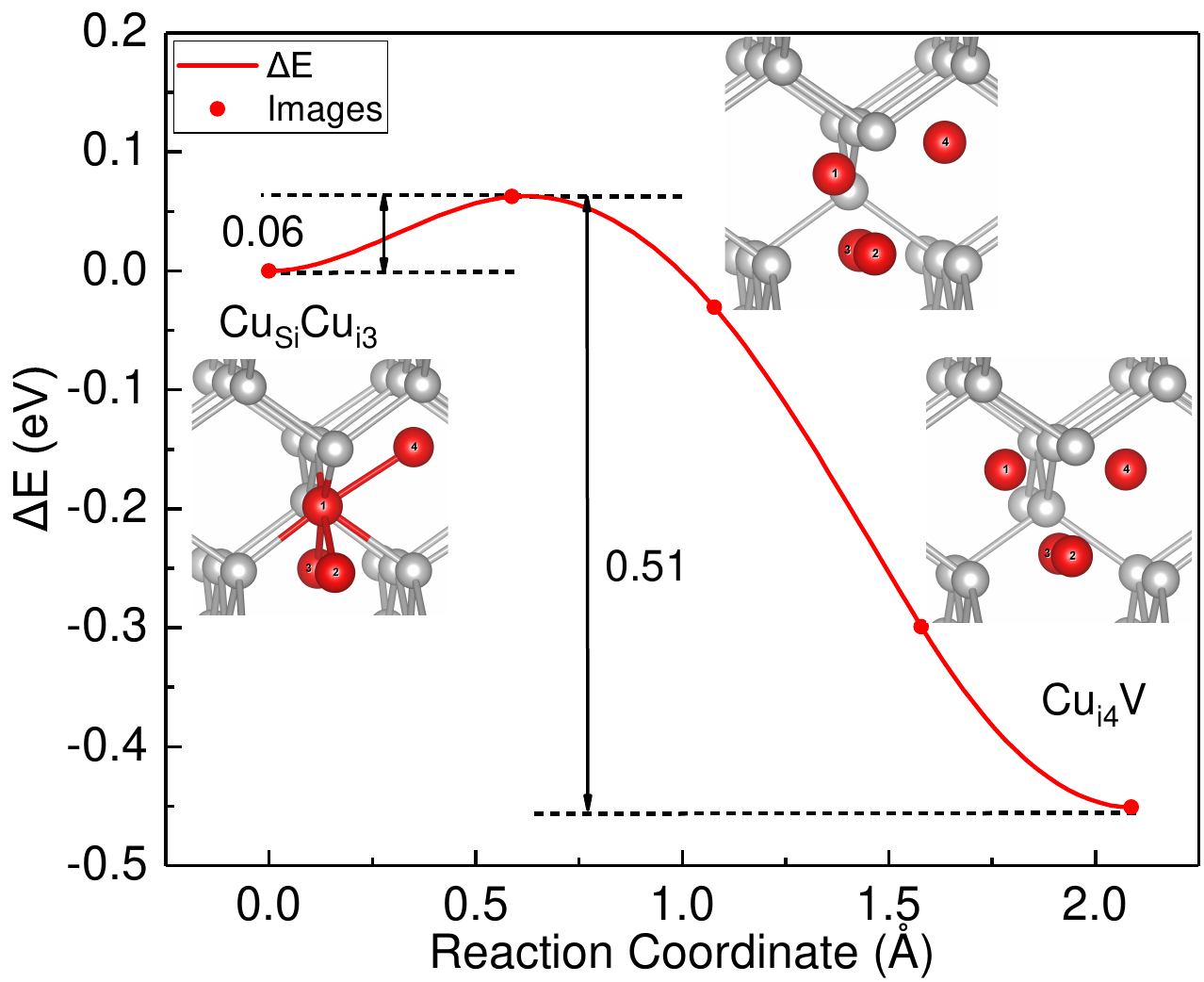}}	
	\subfigure{
		\put(0,175){\textbf{(b)}}
		\label{CuPL_formation}
		\includegraphics[width=0.95\linewidth]{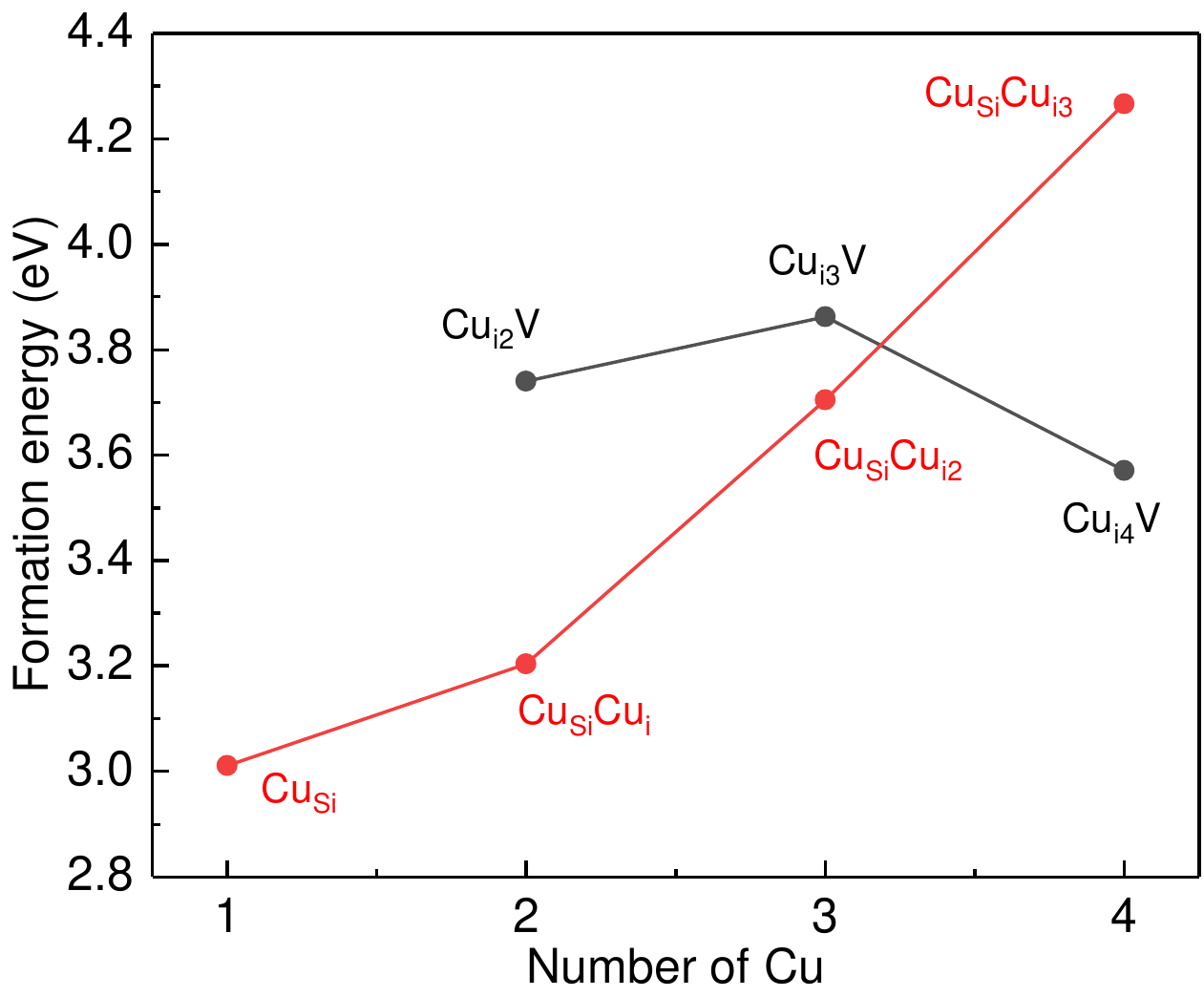}}
	\subfigure{
		\put(0,190){\textbf{(c)}}
		\label{FormE_CusiCui}
		\includegraphics[width=0.95\linewidth]{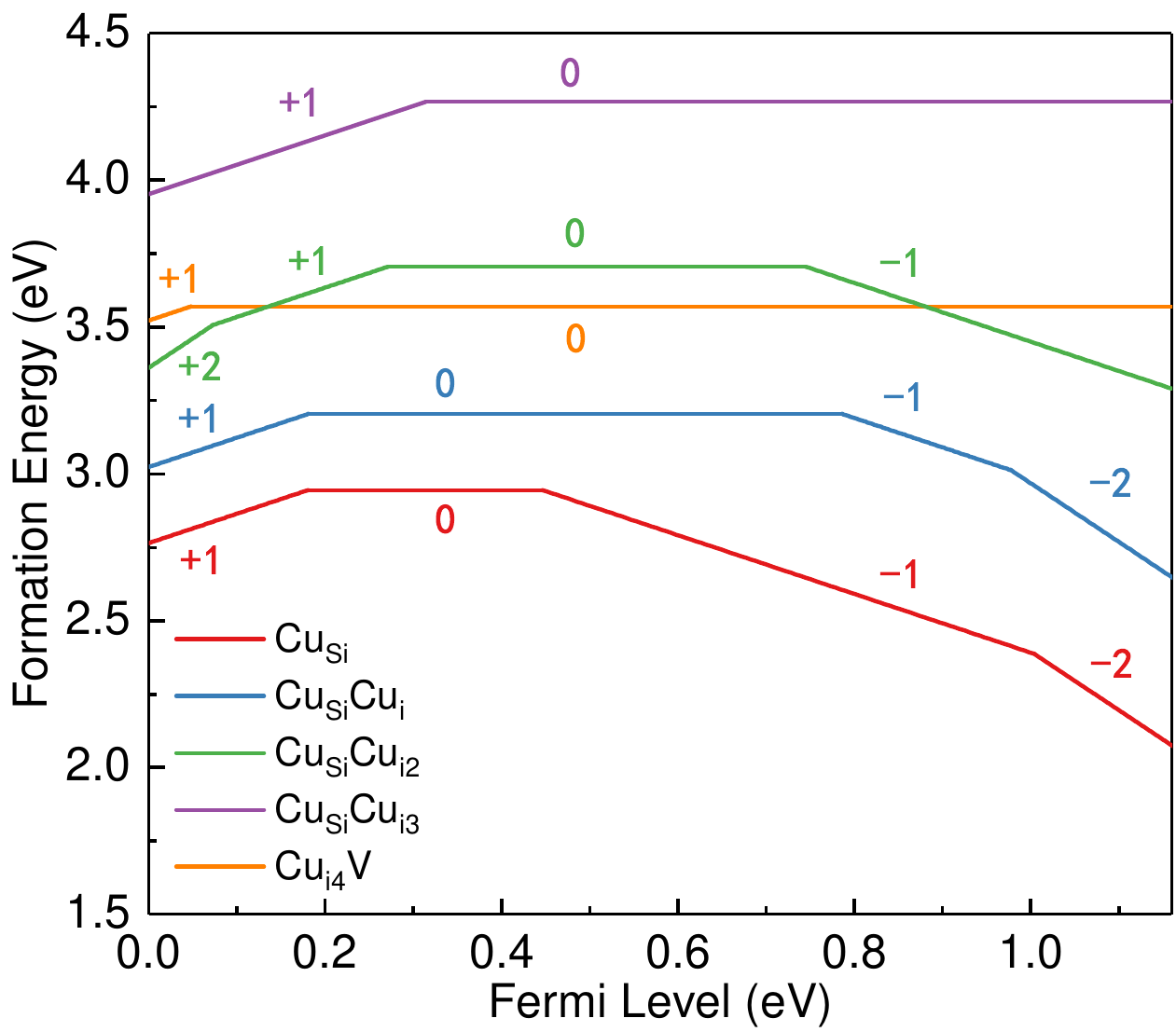}}			
	\caption{(a) Energy barriers between $\mathrm{Cu_{Si}\text{-}Cu_{i3}}$ and $\mathrm{Cu_{i4}V}$. Atomic configurations at initial state ($\mathrm{Cu_{Si}\text{-}Cu_{i3}}$), saddle point and final state ($\mathrm{Cu_{i4}V}$) are shown in the insets. (b) Calculated formation energies of defects associated with $\mathrm{Cu_{PL}}$. (c) Calculated formation energies of $\mathrm{Cu_{Si}\text{-}Cu_{in}}$(n=0, 1, 2, 3) defects.  }
\end{figure}

The configurations of $\mathrm{Cu_{Si}\text{-}Cu_{i3}}$ and $\mathrm{Cu_{i4}V}$ are shown in~\cref{CuPL_NEB}. $\mathrm{Cu_{Si}\text{-}Cu_{i3}}$ has ${C_{3v}}$ symmetry, with all three $\mathrm{Cu_{Si}\text{-}Cu_{i}}$ bonds being 2.36~\textnormal{\AA} at neutral charge state. $\mathrm{Cu_{i4}V}$ has ${T_d}$ symmetry where four interstitial copper surrounds a vacancy. In $\mathrm{Cu_{i4}V^0}$, all $\mathrm{Cu_{i}\text{-}Cu_{i}}$ bond lengths are 2.53~\textnormal{\AA}.

From the perspective of formation energies and transition levels, $\mathrm{Cu_{i4}V}$ is a more plausible candidate. Our calculation shows that $\mathrm{Cu_{Si}\text{-}Cu_{i3}}$ has a $(+1/0)$ level at $E_\text{v}+0.32~\text{eV}$, deviating from the experimental value of $E_\text{v}+0.10~\text{eV}$ (See~\cref{CuPL_level}). Instead, $\mathrm{Cu_{i4}V}$ has a $(+1/0)$ level at $E_\text{v}+0.07~\text{eV}$. Moreover, formation energy of $\mathrm{Cu_{i4}V^0}$ is 0.45 eV lower than $\mathrm{{Cu_{Si}\text{-}Cu_{i3}}^0}$. 

The calculated formation energies of defects associated with $\mathrm{Cu_{PL}}$ are shown~\cref{CuPL_formation}. The formation mechanism of $\mathrm{Cu_{PL}}$ defect can be described as a sequential trapping process of $\mathrm{Cu_i}$. Initially, in Cu-contaminated Si, especially FZ-Si where more vacancies are released during heat treatment~\cite{yarykin2024copperricha}, Cu exists mainly as $\mathrm{Cu_{Si}}$ and $\mathrm{Cu_i}$. $\mathrm{Cu_{Si}^-}$ is likely to trap a diffusing $\mathrm{Cu_{i}^+}$:

\begin{table*}[!htbp]
	\centering
	\renewcommand\arraystretch{1.2}
	\setlength{\tabcolsep}{4pt}
	\caption{Calculated transition levels of defects associated with $\mathrm{Cu_{PL}}$ of this work, along with previously reported DFT results (calc.) and experimental data (exp.). All values are presented in eV relative to $E_\text{v}$. An asterisk (*) means that the acceptor level is originally related to $E_\text{c}$, and recalculated to $E_\text{v}$, assuming a band gap of 1.12 eV.}
	\begin{tabular}{c *{5}{c c c}}
		\toprule
		Transition  & \multicolumn{3}{c}{$\mathrm{Cu_{Si}\text{-}Cu_{i}}$} & \multicolumn{3}{c}{$\mathrm{Cu_{Si}\text{-}Cu_{i2}}$} & \multicolumn{3}{c}{$\mathrm{Cu_{Si}\text{-}Cu_{i3}}$} & \multicolumn{3}{c}{$\mathrm{Cu_{i4}V}$} & \multicolumn{3}{c}{$\mathrm{Cu_{PL}}$} \\
		\cmidrule(lr){2-4} \cmidrule(lr){5-7} \cmidrule(lr){8-10} \cmidrule(lr){11-13} \cmidrule(lr){14-16}
		level & t.w. & calc. & exp. & t.w. & calc. & exp. & t.w. & calc. & exp.& t.w. & calc. & exp. & t.w. & calc. & exp.\\
		\midrule
		(+2/+1) &      &  &  & 0.07 &  &  &  & 0.13$^a$  &  &  &  &  &  &  &  \\
		&      &  &  &  &  &  &  & 0.26$^b$  &  &  &  &  &  &  &  \\
		(+1/0)  & 0.18 & 0.23$^a$ &  & 0.27 & 0.27$^a$ &  & 0.32 & 0.30$^a$ &  & 0.07 & 0.07$^c$ &  &  &  & 0.10$^d$ \\
		&  & 0.18$^b$ &  &  &  &  &  & 0.37$^b$  &  &  &  &  &  &  &  \\
		(0/$-$1)  & 0.78 & 0.55*$^a$ &  & 0.75 & 0.56*$^a$ &  &      & 1.10*$^a$ &  &  &  &  &  &  &  \\
		&  & 0.63*$^b$ &  &  &  &  &  &  &  &  &  &  &  &  &  \\
		($-$1/$-$2) & 0.98 & 1.00*$^a$ &  &  &  &  &  &  &  &  &  &  &  &  &  \\
		&  & 1.00*$^b$ &  &  &  &  &  &  &  &  &  &  &  &  &  \\
		\bottomrule
		\multicolumn{13}{l}{$^{\mathrm{a}}$ Calculated with finite-size correction in a 216-atom supercell. See Ref.~\cite{vincent2020cu}} \\
		\multicolumn{13}{l}{$^{\mathrm{b}}$ Calculated with finite-size correction in a 64-atom supercell. See Ref.~\cite{sharan2017hybridfunctional}} \\
		\multicolumn{13}{l}{$^{\mathrm{c}}$ Calculated with no correction in a 1000-atom supercell. See Ref.~\cite{fujimura2021revisiting}.}\\
		\multicolumn{13}{l}{$^{\mathrm{d}}$ Experimental results by DLTS. See Ref.~\cite{weber1982optical}}
	\end{tabular}
	\label{CuPL_level}
\end{table*}

$\mathrm{Cu_{Si}^-+Cu_{i}^+\rightarrow {Cu_{Si}\text{-}Cu_{i}}^0}$~~~~~~~~~~~~~~~~$E_\text{b}$=1.27 eV 

$\mathrm{Cu_{Si}\text{-}Cu_{i}}$ complex can carry $-1$ charge, allowing it to remain an effective trap for $\mathrm{Cu_i^+}$:

$\mathrm{Cu_{Si}\text{-}Cu_{i}^-+Cu_{i}^+\rightarrow {Cu_{Si}\text{-}Cu_{i2}}^0}$ ~~~~~$E_\text{b}$=1.22 eV 

\noindent A third $\mathrm{Cu_{i}^+}$ can be trapped in the same way:

$\mathrm{{Cu_{Si}\text{-}Cu_{i2}}^-+Cu_{i}^+\rightarrow {Cu_{Si}\text{-}Cu_{i3}}^0}$ ~~~$E_\text{b}$=1.20 eV

Without the assistance of Coulomb attraction, binding energy of a fourth $\mathrm{Cu_{i}^+}$ is much smaller, indicating that the growth of Cu cluster stops at $\mathrm{Cu_{Si}\text{-}Cu_{i3}}$, and the following reaction is unlikely to occur:

$\mathrm{{Cu_{Si}\text{-}Cu_{i3}}^0+Cu_{i}^+\rightarrow {Cu_{Si}\text{-}Cu_{i4}}^+}$ $E_\text{b}$=0.21 eV

In contrast to the case of $\mathrm{H_i}$, the formation energy of $\mathrm{Cu_{Si}\text{-}Cu_{in}}$ complexes increases as additional $\mathrm{Cu_i}$ atoms are introduced around the $\mathrm{Cu_{Si}}$ center (See~\cref{FormE_CusiCui}). This trend suggests that Cu clustering near the substitutional site is energetically less favorable, likely due to the lattice strain introduced by multiple Cu atoms and the relatively weak Cu–Cu interaction in the Si lattice. These contrasting trends highlight the different roles of hydrogen passivation and Cu aggregation in determining the stability of Cu-related defect complexes. Such competition between H passivation and Cu clustering may influence the evolution of Cu defects under hydrogen-rich processing conditions.

~\cref{CuPL_NEB} shows the NEB calculation between $\mathrm{Cu_{i4}V^0}$ and $\mathrm{{Cu_{Si}\text{-}Cu_{i3}}^0}$. The transformation from $\mathrm{Cu_{i4}V^0}$ to $\mathrm{{Cu_{Si}\text{-}Cu_{i3}}^0}$ has a barrier of only 0.06 eV. Therefore, upon annealing at about $400^\circ\mathrm{C}$, a typical condition for Cu in-diffusion experiments~\cite{carvalho2011fourcopper}, $\mathrm{Cu_{Si}\text{-}Cu_{i3}}$ may transform into $\mathrm{Cu_{i4}V}$. The reverse reaction has a barrier of 0.51 eV, which is too high to overcome at room temperature. DLTS signal of $\mathrm{Cu_{PL}}$ disappears and $\mathrm{Cu_{Si}}$ appears after annealing at $250^\circ\mathrm{C}$ to $350^\circ\mathrm{C}$ for 30 minutes~\cite{knack2004copperrelated,vincent2020cu}. This can be explained by the formation energy shown in~\cref{CuPL_formation}. Upon heating, $\mathrm{Cu_{i4}V}$ dissociates with the release of $\mathrm{Cu_i}$. $\mathrm{Cu_{i3}V}$ and $\mathrm{Cu_{i2}V}$ exhibit a relatively high formation energy, making them thermodynamically unstable. They consequently transform into $\mathrm{Cu_{Si}\text{-}Cu_{i2}}$. More $\mathrm{Cu_i}$ dissociate after further thermal treatment, and the last $\mathrm{Cu_i}$ occupies the vacancy, ultimately forming $\mathrm{Cu_{Si}}$. 

Calculated transition levels of defects associated with $\mathrm{Cu_{PL}}$ are listed in~\cref{CuPL_level}. Experiment results for $\mathrm{Cu_{Si}\text{-}Cu_{i}}$ and $\mathrm{Cu_{Si}\text{-}Cu_{i2}}$ are inadequate due to their instability. Our result agrees with previous calculations except for $(0/-1)$ levels of $\mathrm{Cu_{Si}\text{-}Cu_{i}}$ and $\mathrm{Cu_{Si}\text{-}Cu_{i2}}$. In general, $\mathrm{Cu_{i4}V}$ matches experimental observations more than $\mathrm{Cu_{Si}\text{-}Cu_{3i}}$ in terms of both transition level and formation energy, thus being a more likely candidate for $\mathrm{Cu_{PL}}$. 

However, uniaxial stress experiments have shown that $\mathrm{Cu_{PL}}$ center must have $C_{3v}$ symmetry~\cite{weber1982optical}, which differs from $T_d$ symmetry of $\mathrm{Cu_{i4}V}$. It should be noted that the experiment probes an exciton bound to the defect, instead of the isolated defect in its ground state configuration~\cite{weber1982optical}. Therefore, the experimentally inferred symmetry corresponded to the exciton–defect complex. In this context, the discrepancy between $T_d$ and $C_{3v}$ may originate from symmetry lowering in the excited state. Such behavior is reminiscent of Jahn–Teller distortion, as it was proposed in previous $\mathrm{Cu_{PL}}$ study~\cite{fujimura2021revisiting}. Moreover, $\mathrm{Cu_{Si}}$ have been predicted to exhibit Jahn–Teller effect~\cite{latham2006electronica,sharan2017hybridfunctional} (See section 3.1), suggesting that similar effect may also be relevant for $\mathrm{Cu_{i4}V}$ complex. Although the exact excited-state configuration is beyond the scope of this work, it is plausible that $\mathrm{Cu_{i4}V}$ has $C_{3v}$ symmetry in the excited state, which was the observed symmetry in experiment.

In general, our results suggest that the $\mathrm{Cu_{i4}V}$ configuration provides an improved description of the ground state of $\mathrm{Cu_{PL}}$ center in terms of the transition level, while remaining energetically competitive with previously proposed models.  

\section{Conclusion}
Stable configurations, transition levels, and binding energies of Cu-related defects in silicon are systematically investigated using the HSE06 hybrid functional under a unified computational framework. 
Cu isolated defects are first examined, demonstrating that $\mathrm{Cu_i}$ is the most energetically favorable defect in silicon with +1 and 0 charge states within the bandgap, whereas $\mathrm{Cu_{Si}}$ is mainly formed when a diffusing $\mathrm{Cu_i}$ is trapped by a pre-existing vacancy. 
Distinct gettering mechanisms of Cu by B and P are identified. In Cu–B complexes, Cu remains in the interstitial configuration and the binding is relatively weak. In Cu–P complexes, Cu tends to occupy substitutional sites. The gettering effect of phosphorus is stronger than that of boron and increases with P concentration.
Configurations of $\mathrm{Cu_{Si}\text{-}H_{in}}$ complexes with different symmetries are identified and shown to depend on the Fermi level. The calculated transition levels of $\mathrm{Cu_{Si}\text{-}H_{in}}$ agree well with experiments. Up to three hydrogen atoms can bind sequentially to $\mathrm{Cu_{Si}}$, passivating its transition levels and improving its stability energetically. 
Finally, $\mathrm{Cu_{i4}V}$ defect is proposed as a candidate for the $\mathrm{Cu_{PL}}$ center. Compared with the previously suggested $\mathrm{Cu_{Si}\text{-}Cu_{i3}}$ model, $\mathrm{Cu_{i4}V}$ exhibits a lower formation energy, and a transition level closer to experiment. The transformation from $\mathrm{Cu_{Si}\text{-}Cu_{i3}}$ to $\mathrm{Cu_{i4}V}$ is driven by the lattice strain introduced by multiple Cu atoms and the relatively weak Cu–Cu interaction. 
Our calculated formation energies and transition levels exhibit relatively good consistency with experimental observations, and provide a consistent picture about the interaction mechanisms of Cu-related defects in silicon.

\printcredits

\section*{Acknowledgment}
This work was supported by research funds from Shanghai Advanced Silicon Technology Co., Ltd., and also the Natural Science Foundation of Shanghai (Grant No. 23ZR1403300).

\bibliographystyle{elsarticle-num-names} 
\bibliography{Cudefect-refs.bib}

@article{weiske2025adsorption,
  title = {Adsorption energies on extended surfaces with CCSD(T) quality: ethylene on Si(001)},
  author = {Weiske, Hendrik and Pecher, Lisa and Gallo, Alejandro and Irmler, Andreas and Hummel, Felix and Gr{\"u}neis, Andreas and {Tonner-Zech}, Ralf},
  year = 2025,
  journal = {Molecular Physics},
  volume = {0},
  number = {0},
  pages = {e2602647},
  issn = {0026-8976},
  doi = {10.1080/00268976.2025.2602647}
}

@article{straumanis1961perfection,
  title = {Perfection of the Lattice of Dislocation-Free Silicon, Studied by the Lattice-Constant and Density Method},
  author = {Straumanis, M. E. and Borgeaud, P. and James, W. J.},
  year = 1961,
  journal = {J. Appl. Phys.},
  volume = {32},
  number = {7},
  pages = {1382--1384},
  issn = {0021-8979},
  doi = {10.1063/1.1736238}
}

@article{istratov1998dissociation,
  title = {The dissociation energy and the charge state of a copper-pair center in silicon},
  author = {Istratov, A. A. and Hieslmair, H. and Heiser, T. and Flink, C. and Weber, E. R.},
  year = 1998,
  journal = {Appl. Phys. Lett.},
  volume = {72},
  number = {4},
  pages = {474--476},
  issn = {0003-6951},
  doi = {10.1063/1.120790}
}

@article{yarykin2016interstitial,
  title = {{Interstitial carbon in p-type copper-doped silicon}},
  author = {Yarykin, Nikolai and Weber, J{\"o}rg},
  year = 2016,
  journal = {Solid State Phenom.},
  volume = {242},
  pages = {302--307},
  issn = {1662-9779},
  doi = {10.4028/www.scientific.net/SSP.242.302}
}

@article{russo2017dark,
  title = {{Dark current spectroscopy of transition metals in CMOS image sensors}},
  author = {Russo, Felice and Nardone, Giancarlo and Polignano, Maria Luisa and D'Ercole, Angelo and Pennella, Fabrizio and Felice, Massimo Di and Monte, Andrea Del and Matarazzo, Antonio and Moccia, Giuseppe and Polsinelli, Gianpaolo and D'Angelo, Antonio and Liverani, Massimo and Irrera, Fernanda},
  year = 2017,
  journal = {ECS J. Solid State Sci. Technol.},
  volume = {6},
  number = {5},
  pages = {P217},
  issn = {2162-8777},
  doi = {10.1149/2.0101705jss}
}

@article{matsukawa2007diffusion,
  title = {{Diffusion of transition-metal impurities in silicon}},
  author = {Matsukawa, K. and Shirai, K. and Yamaguchi, H. and {Katayama-Yoshida}, H.},
  year = 2007,
  journal = {Physica B: Condensed Matter},
  volume = {401--402},
  pages = {151--154},
  issn = {0921-4526},
  doi = {10.1016/j.physb.2007.08.134}
}

@article{brotherton1987deepa,
  title = {{Deep levels of copper in silicon}},
  author = {Brotherton, S. D. and Ayres, J. R. and Gill, A. and {van Kesteren}, H.W. and Greidanus, F. J. A. M.},
  year = 1987,
  journal = {J. Appl. Phys.},
  volume = {62},
  number = {5},
  pages = {1826--1832},
  issn = {0021-8979},
  doi = {10.1063/1.339564}
}

@article{carvalho2011fourcopper,
  title = {{Four-copper complexes in Si and the Cu-photoluminescence defect: A first-principles study}},
  author = {Carvalho, A. and Backlund, D. J. and Estreicher, S. K.},
  year = 2011,
  journal = {Phys. Rev. B},
  volume = {84},
  number = {15},
  pages = {155322},
  doi = {10.1103/PhysRevB.84.155322}
}

@article{chen2022firstprinciples,
  title = {{First-principles study of copper contamination in silicon semiconductor}},
  author = {Chen, Pei and Li, Yadong and Qin, Fei and An, Tong and Dai, Yanwei and Zhang, Min and Liu, Minghui and Zhang, Lipeng},
  year = 2022,
  journal = {Surf. Interfaces},
  volume = {31},
  pages = {102084},
  issn = {2468-0230},
  doi = {10.1016/j.surfin.2022.102084}
}

@article{estreicher2004firstprinciples,
  title = {{First-principles theory of copper in silicon}},
  author = {Estreicher, Stefan K.},
  year = 2004,
  journal = {Mater. Sci. Semicond. Process.},
  series = {Copper interaction with silicon based materials: a survey},
  volume = {7},
  number = {3},
  pages = {101--111},
  issn = {1369-8001},
  doi = {10.1016/j.mssp.2004.06.004}
}

@article{fujimura2021revisiting,
  title = {{Revisiting the stable structure of the Cu$_4$ complex in silicon}},
  author = {Fujimura, Takayoshi and Shirai, Koun},
  year = 2021,
  journal = {Jpn. J. Appl. Phys.},
  volume = {60},
  number = {2},
  pages = {021001},
  issn = {1347-4065},
  doi = {10.35848/1347-4065/abd495}
}

@article{istratov1997interstitial,
  title = {{Interstitial copper-related center in n-type silicon}},
  author = {Istratov, A. A. and Hieslmair, H. and Flink, C. and Heiser, T. and Weber, E. R.},
  year = 1997,
  journal = {Appl. Phys. Lett.},
  volume = {71},
  number = {16},
  pages = {2349--2351},
  issn = {0003-6951},
  doi = {10.1063/1.120026}
}

@article{istratov1998intrinsica,
  title = {{Intrinsic diffusion coefficient of interstitial copper in silicon}},
  author = {Istratov, Andrei A. and Flink, Christoph and Hieslmair, Henry and Weber, Eicke R. and Heiser, Thomas},
  year = 1998,
  journal = {Phys. Rev. Lett.},
  volume = {81},
  number = {6},
  pages = {1243--1246},
  doi = {10.1103/PhysRevLett.81.1243}
}

@article{istratov2001physics,
  title = {{Physics of copper in silicon}},
  author = {Istratov, Andrei A. and Weber, Eicke R.},
  year = 2001,
  journal = {J. Electrochem. Soc.},
  volume = {149},
  number = {1},
  pages = {G21},
  issn = {1945-7111},
  doi = {10.1149/1.1421348}
}

@article{kaniewski1997hydrogenation,
  title = {{Hydrogenation of copper related deep states in n-type Si containing extended defects}},
  author = {Kaniewski, J. and Kaniewska, M. and Ornoch, L. and Sekiguchi, Takashi and Sumino, Koji},
  year = 1997,
  journal = {Mater. Sci. Forum},
  volume = {258--263},
  pages = {319--324},
  issn = {1662-9752},
  doi = {10.4028/www.scientific.net/MSF.258-263.319}
}

@article{knack2002copperhydrogen,
  title = {{Copper-hydrogen complexes in silicon}},
  author = {Knack, S. and Weber, J. and Lemke, H. and Riemann, H.},
  year = 2002,
  journal = {Phys. Rev. B},
  volume = {65},
  number = {16},
  pages = {165203},
  doi = {10.1103/PhysRevB.65.165203}
}

@article{knack2004copperrelated,
  title = {{Copper-related defects in silicon}},
  author = {Knack, S.},
  year = 2004,
  journal = {Mater. Sci. Semicond. Process.},
  series = {Copper interaction with silicon based materials: a survey},
  volume = {7},
  number = {3},
  pages = {125--141},
  issn = {1369-8001},
  doi = {10.1016/j.mssp.2004.06.002}
}

@article{latham2005passivation,
  title = {{Passivation of copper in silicon by hydrogen}},
  author = {Latham, C. D. and Alatalo, M. and Nieminen, R. M. and Jones, R. and {\"O}berg, S. and Briddon, P. R.},
  year = 2005,
  journal = {Phys. Rev. B},
  volume = {72},
  number = {23},
  pages = {235205},
  doi = {10.1103/PhysRevB.72.235205}
}

@article{latham2006electronica,
  title = {{Electronic structure calculations for substitutional copper and monovacancies in silicon}},
  author = {Latham, C D and Ganchenkova, M and Nieminen, R M and Nicolaysen, S and Alatalo, M and {\"O}berg, S and Briddon, P R},
  year = 2006,
  journal = {Phys. Scr.},
  volume = {2006},
  number = {T126},
  pages = {61},
  issn = {1402-4896},
  doi = {10.1088/0031-8949/2006/T126/014}
}

@article{lowther2010aggregation,
  title = {{Aggregation of interstitial copper atoms in silicon}},
  author = {Lowther, J. E.},
  year = 2010,
  journal = {Mater. Sci. Semicond. Process.},
  volume = {13},
  number = {1},
  pages = {29--33},
  issn = {1369-8001},
  doi = {10.1016/j.mssp.2010.02.003}
}

@article{matsukawa2006gettering,
  title = {{Gettering mechanism of transition metals in silicon calculated from first principles}},
  author = {Matsukawa, Kazuhito and Hattori, Nobuyoshi and Maegawa, Shigeto and Shirai, Koun and {Katayama-Yoshida}, Hiroshi},
  year = 2006,
  journal = {Physica B: Condensed Matter},
  series = {Proceedings of the 23rd International Conference on Defects in Semiconductors},
  volume = {376--377},
  pages = {224--226},
  issn = {0921-4526},
  doi = {10.1016/j.physb.2005.12.059}
}

@inproceedings{nikolai2011formation,
  title = {{Formation of copper-related deep-level centers in irradiated p-type silicon}},
  booktitle = {{Gettering and defect engineering in semiconductor technology XIV}},
  author = {Yarykin, Nikolai and Weber, J{\"o}rg},
  editor = {Jantsch, W. and Schaffler, F.},
  year = 2011,
  volume = {178--179},
  pages = {154-157},
  address = {Durnten-Zurich},
  issn = {1012-0394},
  doi = {10.4028/www.scientific.net/SSP.178-179.154}
}

@article{ozaki2019gettering,
  title = {{Gettering mechanism of copper in n-type silicon wafers}},
  author = {Ozaki, Rie and Torigoe, Kazuhisa and Mizuno, Taisuke and Yamamoto, Kazuhiro},
  year = 2019,
  journal = {Phys. Status Solidi A},
  volume = {216},
  number = {17},
  pages = {1900220},
  issn = {1862-6319},
  doi = {10.1002/pssa.201900220},
  copyright = {\copyright{} 2019 WILEY-VCH Verlag GmbH \& Co. KGaA, Weinheim}
}

@article{sharan2017hybridfunctional,
  title = {{Hybrid-functional calculations of the copper impurity in silicon}},
  author = {Sharan, Abhishek and Gui, Zhigang and Janotti, Anderson},
  year = 2017,
  journal = {Phys. Rev. Appl.},
  volume = {8},
  number = {2},
  pages = {024023},
  doi = {10.1103/PhysRevApplied.8.024023}
}

@article{shirai2005molecular,
  title = {{Molecular dynamics study of fast diffusion of Cu in silicon}},
  author = {Shirai, Koun and Michikita, Toshiyuki and {Katayama-Yoshida}, H.},
  year = 2005,
  journal = {Jpn. J. Appl. Phys.},
  volume = {44},
  pages = {7760--7764},
  doi = {10.1143/JJAP.44.7760}
}

@article{shirai2009new,
  title = {{A new structure of Cu complex in Si and its photoluminescence}},
  author = {Shirai, K and Yamaguchi, H and Yanase, A and {Katayama-Yoshida}, H},
  year = 2009,
  journal = {J. Phys.: Condens. Matter},
  volume = {21},
  number = {6},
  pages = {064249},
  issn = {0953-8984},
  doi = {10.1088/0953-8984/21/6/064249}
}

@article{shirasawa2015useful,
  title = {{Useful database of effective gettering sites for metal impurities in Si wafers with first principles calculation}},
  author = {Shirasawa, Sho and Sueoka, Koji and Yamaguchi, Tadashi and Maekawa, Kazuyoshi},
  year = 2015,
  journal = {ECS J. Solid State Sci. Technol.},
  volume = {4},
  number = {9},
  pages = {P351},
  issn = {2162-8777},
  doi = {10.1149/2.0051509jss}
}

@article{steger2008reduction,
  title = {{Reduction of the linewidths of deep luminescence centers in {$^{28}$Si} reveals fingerprints of the isotope constituents}},
  author = {Steger, M. and Yang, A. and Stavrias, N. and Thewalt, M. L. W. and Riemann, H. and Abrosimov, N. V. and Churbanov, M. F. and Gusev, A. V. and Bulanov, A. D. and Kovalev, I. D. and Kaliteevskii, A. K. and Godisov, O. N. and Becker, P. and Pohl, H.-J.},
  year = 2008,
  journal = {Phys. Rev. Lett.},
  volume = {100},
  number = {17},
  pages = {177402},
  doi = {10.1103/PhysRevLett.100.177402}
}

@article{vandewalle1989theory,
  title = {{Theory of hydrogen diffusion and reactions in crystalline silicon}},
  author = {{Van de Walle}, Chris G.},
  year = 1989,
  journal = {Phys. Rev. B},
  volume = {39},
  number = {15},
  pages = {10791--10808},
  doi = {10.1103/PhysRevB.39.10791}
}

@article{vincent2020cu,
  title = {{The Cu photoluminescence defect and the early stages of Cu precipitation in Si}},
  author = {Vincent, T. M. and Estreicher, S. K. and Weber, J. and Kolkovsky, V. and Yarykin, N.},
  year = 2020,
  journal = {J. Appl. Phys.},
  volume = {127},
  number = {8},
  pages = {085704},
  issn = {0021-8979},
  doi = {10.1063/1.5140456}
}

@article{weber1982optical,
  title = {{Optical properties of copper in silicon: Excitons bound to isoelectronic copper pairs}},
  author = {Weber, J. and Bauch, H. and Sauer, R.},
  year = 1982,
  journal = {Phys. Rev. B},
  volume = {25},
  number = {12},
  pages = {7688--7699},
  doi = {10.1103/PhysRevB.25.7688}
}

@article{west2003copper,
  title = {{Copper interactions with H, O, and the self-interstitial in silicon}},
  author = {West, D. and Estreicher, S. K. and Knack, S. and Weber, J.},
  year = 2003,
  journal = {Phys. Rev. B},
  volume = {68},
  number = {3},
  pages = {035210},
  doi = {10.1103/PhysRevB.68.035210}
}

@article{wright2016firstprinciples,
  title = {{A first-principles model of copper--boron interactions in Si: Implications for the light-induced degradation of solar Si}},
  author = {Wright, E. and Coutinho, J. and {\"O}berg, S. and Torres, V. J. B.},
  year = 2016,
  journal = {J. Phys.: Condens. Matter},
  volume = {29},
  number = {6},
  pages = {065701},
  issn = {0953-8984},
  doi = {10.1088/1361-648X/aa4d78}
}

@incollection{yakimov2019metal,
  title = {{Metal impurities and gettering in crystalline silicon}},
  booktitle = {{Handbook of photovoltaic silicon}},
  author = {Yakimov, Eugene B.},
  year = 2019,
  pages = {1--46},
  doi = {10.1007/978-3-662-52735-1_23-1},
  isbn = {978-3-662-52735-1}
}

@article{yarykin2011copperrelated,
  title = {{Copper-related deep-level centers in irradiated p-type silicon}},
  author = {Yarykin, Nikolai and Weber, J{\"o}rg},
  year = 2011,
  journal = {Phys. Rev. B},
  volume = {83},
  number = {12},
  pages = {125207},
  doi = {10.1103/PhysRevB.83.125207}
}

@article{yarykin2013deep,
  title = {{Deep levels of copper-hydrogen complexes in silicon}},
  author = {Yarykin, Nikolai and Weber, J{\"o}rg},
  year = 2013,
  journal = {Phys. Rev. B},
  volume = {88},
  number = {8},
  pages = {085205},
  doi = {10.1103/PhysRevB.88.085205}
}

@article{yarykin2024copperricha,
  title = {{Copper-rich complexes in irradiated silicon}},
  author = {Yarykin, Nikolai and Weber, J{\"o}rg},
  year = 2024,
  journal = {J. Appl. Phys.},
  volume = {136},
  number = {12},
  issn = {0021-8979},
  doi = {10.1063/5.0232388}
}

@article{freysoldt2011electrostatic,
  title = {{Electrostatic interactions between charged defects in supercells}},
  author = {Freysoldt, Christoph and Neugebauer, J{\"o}rg and {Van de Walle}, Chris G.},
  year = 2011,
  journal = {Phys. Status Solidi B},
  volume = {248},
  number = {5},
  pages = {1067--1076},
  issn = {1521-3951},
  doi = {10.1002/pssb.201046289},
  copyright = {Copyright \copyright{} 2011 WILEY-VCH Verlag GmbH \& Co. KGaA, Weinheim}
}

@article{freysoldt2014firstprinciples,
  title = {{First-principles calculations for point defects in solids}},
  author = {Freysoldt, Christoph and Grabowski, Blazej and Hickel, Tilmann and Neugebauer, J{\"o}rg and Kresse, Georg and Janotti, Anderson and {Van de Walle}, Chris G.},
  year = 2014,
  journal = {Rev. Mod. Phys.},
  volume = {86},
  number = {1},
  pages = {253--305},
  doi = {10.1103/RevModPhys.86.253}
}

@article{henkelmanClimbingImageNudged2000,
  title = {{A climbing image nudged elastic band method for finding saddle points and minimum energy paths}},
  author = {Henkelman, Graeme and Uberuaga, Blas P. and J{\'o}nsson, Hannes},
  year = 2000,
  journal = {J. Chem. Phys.},
  volume = {113},
  number = {22},
  pages = {9901--9904},
  issn = {0021-9606},
  doi = {10.1063/1.1329672}
}

@article{kresse1996efficient,
  title = {{Efficient iterative schemes for ab initio total-energy calculations using a plane-wave basis set}},
  author = {Kresse, G. and Furthm{\"u}ller, J.},
  year = 1996,
  journal = {Phys. Rev. B},
  volume = {54},
  number = {16},
  pages = {11169--11186},
  doi = {10.1103/PhysRevB.54.11169}
}

@article{blochl1994projector,
  title = {{Projector augmented-wave method}},
  author = {Bl{\"o}chl, P. E.},
  year = 1994,
  journal = {Phys. Rev. B},
  volume = {50},
  number = {24},
  pages = {17953--17979},
  doi = {10.1103/PhysRevB.50.17953}
}

@article{monkhorst1976special,
  title = {{Special points for Brillouin-zone integrations}},
  author = {Monkhorst, Hendrik J. and Pack, James D.},
  year = 1976,
  journal = {Phys. Rev. B},
  volume = {13},
  number = {12},
  pages = {5188--5192},
  doi = {10.1103/PhysRevB.13.5188}
}

@book{claeys2018metal,
  title = {{Metal impurities in silicon- and germanium-based technologies: Origin, characterization, control, and device impact}},
  author = {Claeys, Cor and Simoen, Eddy},
  year = 2018,
  series = {Springer series in materials science},
  volume = {270},
  address = {Cham},
  doi = {10.1007/978-3-319-93925-4},
  copyright = {http://www.springer.com/tdm},
  isbn = {978-3-319-93924-7 978-3-319-93925-4}
}

@article{honda1984breakdown,
   author = {{K. Honda, A. Ohsawa, N. Toyokura}},
   title = {{Breakdown in silicon oxides—correlation with Cu precipitates}},
   journal = {Appl. Phys. Lett.},
   volume = {45},
   number = {3},
   pages = {270-271},
   year = {1984},
   type = {Journal Article}
}

@article{bludau1974temperaturea,
  title = {Temperature dependence of the band gap of silicon},
  author = {{W. Bludau, A. Onton, W. Heinke}},
  year = 1974,
  journal = {J. Appl. Phys.},
  volume = {45},
  number = {4},
  pages = {1846--1848},
  issn = {0021-8979},
  doi = {10.1063/1.1663501}
}

@article{lindroos2016review,
  title = {{Review of light-induced degradation in crystalline silicon solar cells}},
  author = {Lindroos, Jeanette and Savin, Hele},
  year = 2016,
  journal = {Sol. Energy Mater. Sol. Cells},
  volume = {147},
  pages = {115--126},
  issn = {0927-0248},
  doi = {10.1016/j.solmat.2015.11.047}
}

@article{bystrom2024nonlocal,
  title = {{Nonlocal machine-learned exchange functional for molecules and solids}},
  author = {Bystrom, Kyle and Kozinsky, Boris},
  year = 2024,
  journal = {Phys. Rev. B},
  volume = {110},
  number = {7},
  pages = {075130},
  doi = {10.1103/PhysRevB.110.075130}
}

@article{lee2022transition,
  title = {{Transition metal impurities in silicon: Computational search for a semiconductor qubit}},
  author = {Lee, Cheng-Wei and Singh, Meenakshi and Tamboli, Adele C. and Stevanovi{\'c}, Vladan},
  year = 2022,
  journal = {npj Comput. Mater.},
  volume = {8},
  number = {1},
  pages = {1--11},
  issn = {2057-3960},
  doi = {10.1038/s41524-022-00862-z},
  copyright = {2022 The Author(s)}
}

@book{drabold2007theory,
  title = {Theory of defects in semiconductors},
  editor = {Drabold, David A. and Estreicher, Stefan K. and Ascheron, Claus E. and K{\"o}lsch, Hans J. and Duhm, Adelheid H.},
  year = 2007,
  series = {Topics in applied physics},
  volume = {104},
  address = {Berlin, Heidelberg},
  doi = {10.1007/11690320},
  copyright = {http://www.springer.com/tdm},
  isbn = {978-3-540-33400-2 978-3-540-33401-9}
}

@article{hara1993hydrogen,
  title = {{Hydrogen effects on oxygen precipitation in Czochralski silicon crystals}},
  author = {Hara, Akito and Aoki, Masaki and Fukuda, Tetsuo and Ohsawa, Akira},
  year = 1993,
  journal = {J. Appl. Phys.},
  volume = {74},
  number = {2},
  pages = {913--916},
  issn = {0021-8979},
  doi = {10.1063/1.354858}
}

@article{jiang2003hydrogenation,
  title = {{Hydrogenation of Si from SiN$_x$(H) films: Characterization of H introduced into the Si}},
  author = {Jiang, Fan and Stavola, Michael and Rohatgi, A. and Kim, D. and Holt, J. and Atwater, H. and Kalejs, J.},
  year = 2003,
  journal = {Appl. Phys. Lett.},
  volume = {83},
  number = {5},
  pages = {931--933},
  issn = {0003-6951},
  doi = {10.1063/1.1598643}
}

@article{nickel2023hydrogen,
  title = {{Hydrogen incorporation in semiconductors}},
  author = {Nickel, Norbert H.},
  year = 2023,
  journal = {Phys. Status Solidi B},
  volume = {260},
  number = {10},
  pages = {2300309},
  issn = {1521-3951},
  doi = {10.1002/pssb.202300309},
  copyright = {\copyright{} 2023 The Authors. physica status solidi (b) basic solid state physics published by Wiley-VCH GmbH}
}

@article{duo2001evolution,
  title = {{Evolution of hydrogen and helium co-implanted single-crystal silicon during annealing}},
  author = {Duo, Xinzhong and Liu, Weili and Zhang, Miao and Wang, Lianwei and Lin, Chenglu and Okuyama, M. and Noda, M. and Cheung, Wing-Yiu and Wong, S. P. and Chu, Paul K. and Hu, Peigang and Wang, S. X. and Wang, L. M.},
  year = 2001,
  journal = {J. Appl. Phys.},
  volume = {90},
  number = {8},
  pages = {3780--3786},
  issn = {0021-8979},
  doi = {10.1063/1.1389478}
}

@article{yarykin2021copper,
  title = {{Copper complexes with carbon-related radiation defects in silicon}},
  author = {Yarykin, Nikolai and Weber, J{\"o}rg and Lastovskii, Stanislau and Gusakov, Vasilii},
  year = 2021,
  journal = {Phys. Status Solidi A},
  volume = {218},
  number = {23},
  pages = {2100141},
  issn = {1862-6319},
  doi = {10.1002/pssa.202100141},
  copyright = {\copyright{} 2021 Wiley-VCH GmbH}
}

@article{markevich2008radiationinduced,
  title = {{Radiation-induced defect reactions in Cz-Si crystals contaminated with Cu}},
  author = {Markevich, Vladimir P. and Peaker, Anthony R. and Medvedeva, I. F. and Gusakov, Vasilii E. and Murin, L. I. and Svensson, Bengt Gunnar},
  year = 2008,
  journal = {Solid State Phenom.},
  volume = {131--133},
  pages = {363--368},
  issn = {1662-9779},
  doi = {10.4028/www.scientific.net/SSP.131-133.363}
}

@article{coletti2011impact,
  title = {{Impact of metal contamination in silicon solar cells}},
  author = {Coletti, Gianluca and Bronsveld, Paula C. P. and Hahn, Giso and Warta, Wilhelm and Macdonald, Daniel and Ceccaroli, Bruno and Wambach, Karsten and Le Quang, Nam and Fernandez, Juan M.},
  year = 2011,
  journal = {Adv. Funct. Mater.},
  volume = {21},
  number = {5},
  pages = {879--890},
  issn = {1616-3028},
  doi = {10.1002/adfm.201000849},
  copyright = {Copyright \copyright{} 2011 WILEY-VCH Verlag GmbH \& Co. KGaA, Weinheim}
}

@article{hiramoto1989degradation,
  title = {{Degradation of gate oxide integrity by metal impurities}},
  author = {Hiramoto, Kazuo and Sano, Masakazu and Sadamitsu, Shinsuke and Fujino, Nobukatsu},
  year = 1989,
  journal = {Jpn. J. Appl. Phys.},
  volume = {28},
  number = {12A},
  pages = {L2109},
  issn = {1347-4065},
  doi = {10.1143/JJAP.28.L2109}
}

@article{istratov1998electrical,
  title = {{Electrical and recombination properties of copper-silicide precipitates in silicon}},
  author = {Istratov, A. A. and Hedemann, H. and Seibt, M. and Vyvenko, O. F. and Schr{\"o}ter, W. and Heiser, T. and Flink, C. and Hieslmair, H. and Weber, E. R.},
  year = 1998,
  journal = {J. Electrochem. Soc.},
  volume = {145},
  number = {11},
  pages = {3889},
  issn = {1945-7111},
  doi = {10.1149/1.1838889}
}

@article{heyd2006erratum,
  title = {{Erratum: ``Hybrid functionals based on a screened Coulomb potential'' [J. Chem. Phys. 118, 8207 (2003)]}},
  author = {Heyd, Jochen and Scuseria, Gustavo E. and Ernzerhof, Matthias},
  year = 2006,
  journal = {J. Chem. Phys.},
  volume = {124},
  number = {21},
  pages = {219906},
  issn = {0021-9606},
  doi = {10.1063/1.2204597}
}


\end{document}